\documentclass[prb,superscriptaddress,twocolumn,floatfix,notitlepage,citeautoscript]{revtex4-2}
\usepackage[T2A]{fontenc}
\usepackage{mmap}
\usepackage{amssymb}
\usepackage{amsmath}
\usepackage{color}
\usepackage[usenames,dvipsnames]{xcolor}
\usepackage{hyperref}
\usepackage{textcomp}
\hypersetup{colorlinks=true,linkcolor=blue,urlcolor=blue,citecolor=blue}
\usepackage[normalem]{ulem}
\usepackage{multirow}
\usepackage{graphicx}

\newcommand{\vor}[1]{{\color{black}#1}}
\newcommand{\gru}[1]{{\color{black}#1}}
\usepackage{array}
\newcolumntype{P}[1]{>{\centering\arraybackslash}p{#1}}

\begin{document}

\title{Keldysh field theory of spin- and valley-distinguished polariton nonlinearities in transition-metal dichalcogenide monolayers}

\author{Anna M.~Grudinina}
\affiliation{Abrikosov Center for Theoretical Physics, Moscow Center for Advanced Studies, 141701 Moscow, Russia}
\affiliation{National Research Nuclear University MEPhI (Moscow Engineering Physics Institute), Kashirskoe shosse 31, 115409 Moscow, Russia}

\author{Nina~S.~Voronova}
\email{nsvoronova@mephi.ru}
\affiliation{National Research Nuclear University MEPhI (Moscow Engineering Physics Institute), Kashirskoe shosse 31, 115409 Moscow, Russia}
\affiliation{Russian Quantum Center, Skolkovo IC, Bolshoy boulevard 30 bld. 1, 121205 Moscow, Russia}

\begin{abstract}
Electrons in transition-metal dichalcogenides (TMDs) possess valley and spin degrees of freedom, which leads to rich exciton and exciton-polariton physics with nontrivial scattering dynamics and enhanced nonlinearities, presenting a key mechanism for photonic devices. Yet, existing descriptions of bosonization and polariton interactions in TMD-based systems overlook the valley degree of freedom as well as the various particles' spins combinations. In this work, we derive a nonequilibrium field-theory approach in the path integral formalism that allows to track all the polariton nonlinearities in the strong coupling regime. We demonstrate that, when all the bright and dark exciton species are considered, the TMD monolayer-based polariton systems feature sixteen different nonlinear contributions due to interactions and even more saturation-related terms. Strikingly, while the interactions of excitons within one valley are overall dominant, we show that 
\gru{the interaction constant corresponding to the 
\vor{scattering of} spin-dark \vor{and bright} excitons 
is much larger than that for bright exciton\vor{s alone,} indicating 
\vor{a} possibly} 
\vor{much higher contribution 
to the blueshift from spin-dark excitons 
than 
from bright intravalley excitons.}
\end{abstract}

\maketitle

\section{Introduction}

Exciton-polaritons are quasiparticles resulting from strong coupling between electronic excitations in semiconductors (excitons) and photons~\cite{QFL}. Polaritons have been studied in both conventional semiconductor quantum-well systems and alternative materials---organic semiconductors~\cite{kena-cohen, whittaker}, perovskites~\cite{brehier, xiong_perovskite}, and two-dimensional transition metal dichalcogenides~\cite{menon2015, luo}. The latter allow for the formation of robust excitons at room temperature, which makes these materials promising for the development of polaritonic devices~\cite{polariton_devices}.  
Being hybrid quasiparticles, polaritons simultaneously inherit strong nonlinearities  
from their excitonic component and a very small effective mass  
from the photonic counterpart. 
Polariton nonlinearity plays a crucial role in various many-body phenomena such as Bose-Einstein condensation~\cite{kasprzak}, superfluidity~\cite{amo, lerario}, vortices formation~\cite{lagoudakis}, Josephson oscillatoions~\cite{lagoudakis_JJ, abbrachi}, etc.
Enhancement of interactions may lead to the polariton blockade regime~\cite{verger}, when the interaction strength between polaritons is strong enough to cause the blueshift of the polariton resonance to be larger than the linewidth, which provides the generation of single-polariton states and induces single-particle quantum effects in polariton systems~\cite{polariton_devices,liew_opinion}.

Theoretically, the polariton nonlinearity can be rigorously derived via considering a many-body polariton system as an electron-hole-photon mixture, within the microscopic treatment that is referred to as the bosonization problem~\cite{kira,combescot,  yamamoto, schwendimann, binder, glazov_bosonization, prb110}. 
Therein, for polariton systems in the low-density limit (when $n_{\rm ex} a_B^2\ll1$, where $n_{\rm ex}$ is the exciton density and $a_B$ the exciton Bohr radius), two types of nonlinearities arise: the exciton-interaction nonlinearity which occurs due to electrostatic repulsion, and the so-called saturation---purely polaritonic effect---which is often  related to the exciton phase-space filling~\cite{schmitt-rink}, and consists of the exciton-mediated exciton-photon interconversion (or, vice versa, the photon-mediated exciton interaction). Polariton nonlinearities are hence often characterized using the interaction and saturation constants, $g_{\rm ex}$ and $g_{\rm sat}$, which are commonly introduced in the $1s$ exciton limit at $T=0$ as the exciton wavefunctions overlap integrals~\cite{yamamoto, schwendimann, binder, glazov_bosonization}.
Importantly, different theoretical approaches to the calculation of these constants within the first Born approximation give the same expressions. Experimentally, these constants can be estimated from the condensate blueshift. 
The saturation that arises naturally in bosonization treatments as a small correction to the exciton interaction in the `rigid exciton' limit~\cite{prb110, malic}, in experiments is sometimes controversially argued to be the dominant mechanism of nonlinearity in polariton systems based both on quantum wells~\cite{richard, richard2026} and transition-metal dichalcogenide (TMD) monolayers~\cite{menon}. 
 
Transition-metal dichalcogenides, semiconductors that can be made atomically thin and are thus referred to as two-dimensional (2D) materials, have recently enjoyed a significant advance of fundamental research and 
are promising for applications~\cite{luo, glazov_review,mak}. The integration of TMD monolayers or heterostructures in microcavities and the achievement of the strong-coupling regime is particularly important, due to robustness of TMD polaritons from cryogenic up to room temperatures. 
Nonlinearities in TMD-based polariton systems are expected to be increased due to nonhydrogenic nature of 2D excitons~\cite{ menon, shahnazaryan2017,stepanov}: their electrostatic interactions are described by the Rytova-Keldysh potential~\cite{keldysh_int}, and the enhancement of polariton interactions by the light-matter coupling in TMDs has also been predicted~\cite{parish}. Experimental reports of $g_{\rm ex}$ and $g_{\rm sat}$ in TMD-based systems deviate from each other~\cite{ menon, stepanov, zhang, zhao}, which can possibly be attributed to the presence of dark excitons~\cite{prb110}---the exciton species that cannot couple to light due to the spin mismatch.
At the same time, apart from spins, electrons and holes in TMD materials possess also the valley degree of freedom, so that numerous exciton fields can be excited and, consequently, may also result in a larger nonlinearity, 
which has not been studied 
in the existing approaches.
This work addresses the open question whether the direct consideration of both valley and spin degrees of freedom may potentially contribute to nonlinearities in TMD-based systems.

In our earlier work, the equilibrium path integral technique was applied to the bosonization problem in exciton-polariton systems in general~\cite{prb110}. 
Here, we extend our consideration to non-equilibrium, using a similar approach to that of 
Ref.~\cite{prl132}, taking into account inherent to TMD materials valleys and spins, and consider the nonlinear terms of the theory. We discuss both the exciton-interaction and saturation nonlinearities in TMD-based polariton systems. In our consideration, we focus on the specific case of WS$_2$ monolayer, since due to the large exciton binding energy and nonlinearity, WS$_2$-based polariton systems have been subject of intensive research. In particular, interactions of WS$_2$-based polaritons were experimentally measured in various configurations, including in encapsulated and suspended monolayers~\cite{polimeno}, systems with artificial potential landscape~\cite{landscape_confinement}, bilayers with varying interlayer distance~\cite{zhao_phonon}, as well as superlattices~\cite{menon,zhao}. For the theoretical estimates, however, none of these studies have taken into account the whole variety of the exciton species arising in the material.
We note that results of the current work are also valid for other tungsten-based TMD monolayers that have the same spin and valley structure, whereas for MoX$_2$, X$=$\{S, Se, Te\}, the derived expressions are applicable with $K\leftrightarrow K'$.

The paper is organized as follows. Section~\ref{sec2} is devoted to the bosonization problem for TMD-based polariton systems, with the whole Zoo of valleys and spins taken into account. Section~\ref{sec3} addresses the inter- and intra-valley exciton interactions, including those assisted by light, and contributions from interactions between different types of exciton fields to polariton nonlinearities. In Section~\ref{sec4} we discuss the case of non-negligible losses, as the developed nonequilibrium theory allows to rigorously take them into account.
Section~\ref{sec5} summarizes our results. The main text is supplemented with Appendix~\ref{appA} containing the description of the Keldysh nonequlibrium approach, and Appendix~\ref{appD} dedicated to the calculation details of the momentum-bright intravalley nonlinearities.\\

\section{Bosonization in TMD-based polariton systems}\label{sec2}

We consider the system within the path integral approach and start with the electron-hole-photon action in terms of the fields of electrons $\Psi_{\!i\lambda}^{\sigma}$ in the band $i$ and valley $\lambda$, with spin projection $\sigma$, effective masses $m_{i\lambda}^{\sigma}$ and energy dispersions $\varepsilon_{\!i \lambda}^{\sigma}({\bf k}) = \pm \left(E_{\rm g}/2 + \hbar^2 {\bf k}^2/2m_{i\lambda}^{\sigma} + \delta_{i\lambda}^{\sigma}\right)$. The field $\Psi^{\sigma}_{\rm ph}$ describes photons with polarization $\sigma$ in a microcavity, with the effective mass $m_{\rm ph}$ and  dispersion $E_{\rm ph}({\bf k}) = E_{\rm ph}^0 + \hbar^2 {\bf k}^2/2 m_{\rm ph}$. The origin of energy is taken in the middle of the bandgap of the width $E_{\rm g}$, $E_{\rm ph}^0$ is the cavity resonance, and $\delta_{i\lambda}^{\sigma}$ corresponds to the spin splitting in the $i$-th band. For simplicity, we restrict ourselves to the case $T=0$. The action within the non-equilibrium approach reads
\begin{widetext}
\begin{multline}\label{action_e-h-ph_initial}
    \mathcal{S}
    =  \mathcal{S}_{b} + \mathcal{S}_{sb} + \!\int \!\! d{\bf r} \!\!\int_{\mathcal{C}} \!\!d t
  \left[\sum_{ \sigma}  \overline{\Psi}_{\rm ph}^{ \sigma}(x)(i\partial_{t} - E_{\rm ph }({\bf\hat{k}})) \Psi_{\rm ph}^{ \sigma}(x) 
     \right.  
        +  \sum_{i, \sigma, \lambda}  \overline{\Psi}_{i\lambda}^{ \sigma}(x) (i \partial_{t} \!-\! \varepsilon_{\!i \lambda}^{\sigma}({\bf\hat{k}})){\Psi}_{i\lambda}^{\sigma}(x) \\
         - g_{\rm R} \sum_{\sigma, \lambda}  \left(\overline{\Psi}_{\rm ph}^{\sigma}(x)\Psi_{\!c \lambda}^{\sigma}(x)\overline{\Psi}_{\!v\lambda}^{\sigma}(x) + {\rm c.c.} \!\right) \!\biggr] 
        - \frac{1}{2} \!\sum\limits_{i,j}\sum\limits_{\substack{\sigma,\sigma'\\\lambda,\lambda'}}
        \int \!\! d{\bf r}d{\bf r}^\prime \!\! \int_{\mathcal{C}} \! dt  \, V({\bf r}-{\bf r}^\prime) \overline{\Psi}_{\!i \lambda}^{\sigma}(x)\Psi_{\!i \lambda}^{\sigma}(x) \overline{\Psi}_{\!j \lambda'}^{\sigma'}(x^\prime) \Psi_{\!j\lambda'}^{\sigma'}   (x^\prime),   
\end{multline}
\end{widetext}
where $\sigma = \{\uparrow, \downarrow\}$ are spin indices, $i,j=\{c,v\}$ denote conduction ($c$) and valence ($v$) bands, $\lambda = \{K, K^{\prime}\}$ are the valley indices; $x \equiv ({\bf r},t)$, ${\hat{\bf k}}$ is the momentum operator, $V({\bf r-r^\prime})$ represents the potential of interaction between charged particles. 
Finally, $g_{\rm R}$ is the amplitude of light-matter coupling (electron-hole annihilation with photon creation and vice versa). We note that in the expression above, the time evolution is defined on the closed time contour $\mathcal{C}$. The dissipation is taken into account by means of the terms $\mathcal{S}_{b}$ and $\mathcal{S}_{sb}$ corresponding to the bath and the system-bath interactions, respectively (see Appendix~\ref{appA} for details). We assume that photons of polarization $\pm1$ are converted into electrons and holes with the total spin projections $\uparrow,\downarrow$ in the same valley, due to the spin conservation and the fact that for electrons in different valleys the momentum conservation is not satisfied, as it would require the momentum of photons to be approximately $|{\bf K} - {\bf K'}|\sim 1/a$, i.e. lying outside the light cone
($a$ is the characteristic scale of the lattice vector; for TMDs, $a\sim \text{\AA}$). In Eq.~(\ref{action_e-h-ph_initial}), we do not take into account the electron-hole exchange interaction between electrons and holes in different valleys~\cite{valley_coh1}. 

Considering the exciton pairing channel in the low-density limit $n_{\rm ex} a_B^2 \ll 1$, one arrives at 16 exciton fields $\Delta^{\sigma \sigma'}_{\lambda \lambda'}(k_1, k_2) = \Psi_{c \lambda}^{\sigma}(k_1)\overline{\Psi}_{v \lambda'}^{\sigma'}(k_2)$, where the four-vector notation is used: $k_{1,2}\equiv(\omega_{1,2}, {\bf k}_{1,2})$. The procedure presented in Ref.~\cite{prb110} for the quasi-equilibrium case~\cite{comment1} allows to obtain the action in terms of the exciton and photon fields. Since the case considered here treatment-wise does not differ from the previously treated cases, we provide the derivation  details for the exciton-photon action with nonlinearities in Appendix~\ref{appA}. In the low-temperature dissipationless limit, the resulting Wannier-like equation in the momentum-frequency representation for the exciton field $\hat\Delta^{\sigma \sigma'}_{\lambda \lambda'}$ can be written as follows:
\begin{subequations}\label{Wannier}
\begin{align}
    &\hspace{-10pt}\left[\varepsilon_{c\lambda}^{\sigma}({\bf k}_1\!) \!-\! \varepsilon_{v\lambda'}^{\sigma'}({\bf k}_2\!) \!-\! \Omega\right] \! \hat\Delta^{\sigma \sigma'}_{\lambda \lambda'}({\bf k}_1, {\bf k}_2, \Omega) \!\nonumber\\&\hspace{10pt}= \!
    \sum_{{\bf k}_3} V({\bf k}_1 \!-\! {\bf k}_3) \hat\Delta^{\sigma \sigma'}_{\lambda \lambda'}({\bf k}_3,{\bf k}_3 \!+\! {\bf k}_2 \!-\! {\bf k}_1, \Omega) \nonumber
    \\[-10pt]& \hspace{60pt}- g_{R} \hat\Psi_{\rm ph}^{ \sigma}({\bf k}_1 \!-\! {\bf k}_2, \Omega) \delta_{\sigma \sigma'} \delta_{\lambda \lambda'},\\
    & 
 \hspace{-10pt}
\Bigl[E_{\rm ph}({\bf k}_1 - {\bf k}_2)- \Omega \Bigr]\hat\Psi_{\rm ph}^{\sigma}({\bf k}_1 - {\bf k}_2,\Omega)\nonumber\\[-2pt] & \hspace{60pt} + g_{\rm R}\sum_{{\bf k}_2}\hat\Delta^{\sigma \sigma}_{\lambda\lambda}({\bf k}_1,{\bf k}_2, \Omega)=0,
\end{align}
\end{subequations}
where the hatted quantities correspond to spinors $(cl, q)$ in the Keldysh space for bosonic fields. 
The $1s$ exciton limit allows for the separation of variables:
$$\Delta^{\sigma\sigma'}_{\lambda\lambda'} \! ({\bf k}_1, {\bf k}_2, \Omega) \!=\! \chi^{\sigma \sigma'}_{\lambda\lambda'} \! \Bigl(\!\tfrac{m_{c\lambda}^{\sigma} {\bf k}_1 \!+ m_{v\lambda'}^{\sigma'} {\bf k}_2 }{m_{c\lambda}^{\sigma} + m_{v\lambda'}^{\sigma'}}\!\Bigr) \tilde{\Delta}^{\sigma\sigma'}_{\lambda\lambda'}({\bf k}_1 \!-\! {\bf k}_2, \Omega).$$
Note that the exciton wavefunction $\chi^{\sigma \sigma'}_{\lambda\lambda'}({\bf p})$ depends on valley and spin indices.

\begin{figure}[t!]
    \centering
    \includegraphics[width=0.85\linewidth]{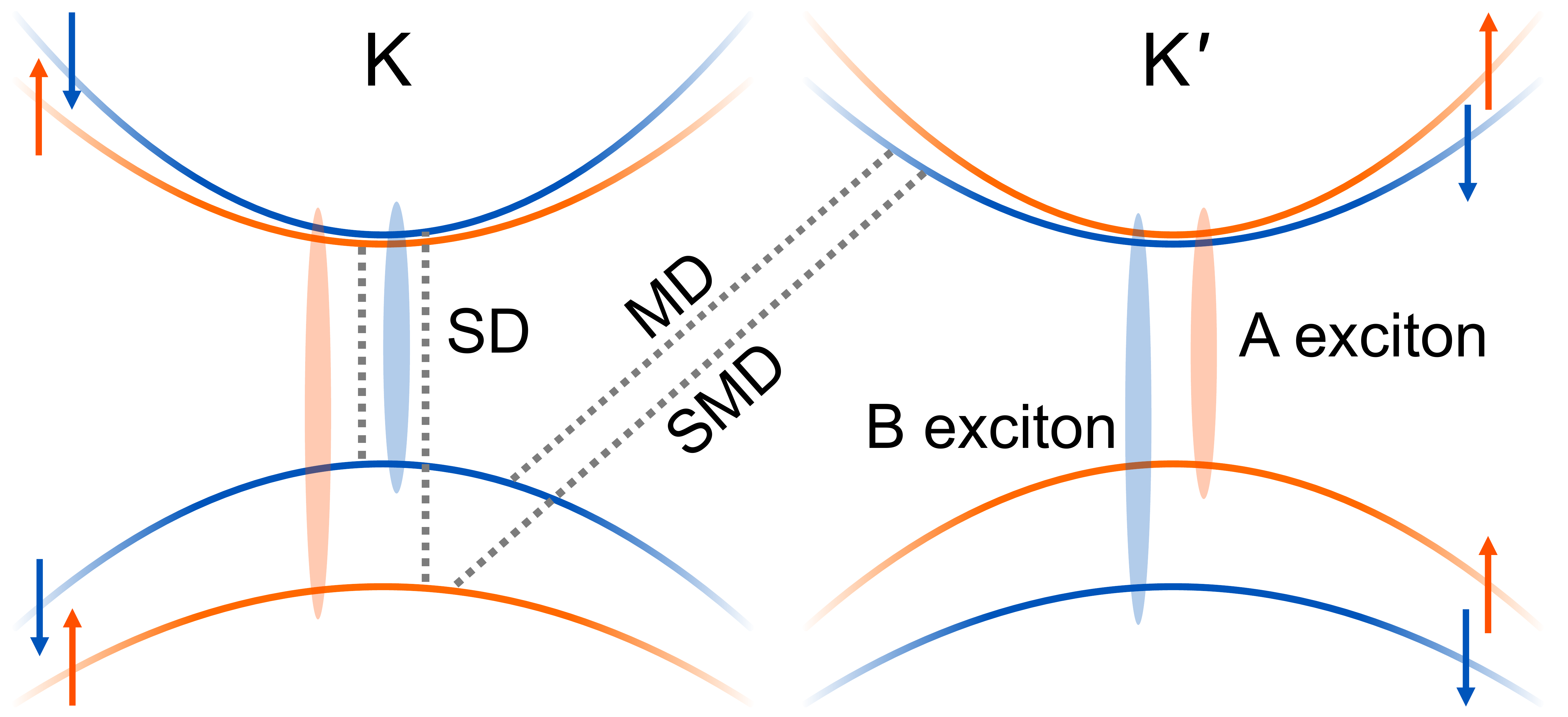}
    \caption{\small Schematic illustration of the electronic band structure near the $K$ and $K'$ points in tungsten-based TMDs. The blue and orange colors correspond to spin-down and spin-up states, respectively.
    The ellipses of respective colors illustrate optical transitions for bright excitons in $K$ and $K'$ valleys, the dashed gray lines correspond to dark fields which are: intravalley spin-dark (SD), intervalley momentum-dark (MD), and spin- and momentum-dark (SMD). 
    }
    \label{fig1}
\end{figure}

From Eqs.~(\ref{Wannier}), one sees that excitons with respect to their spins and valleys can be characterized in terms of being bright---allowing for coupling with photons---or dark, not coupled to photons but affecting the optoelectronic properties through their interaction with bright excitons. A schematic illustration of the formation of dark and bright excitons is shown in Fig.~\ref{fig1}. One can distinguish spin-bright excitons with $\pm 1$ spin projections and spin-dark excitons with the spin projection  $0$, as well as momentum-bright excitons formed by the electrons in the same valley, and momentum-dark excitons consisting of electrons with different valley indices. 
It is important to note that although the spin and valley indices $\sigma$ and $\lambda$ seem to enter the Eqs.~(\ref{Wannier}) in a similar way, their influence is different as the valley degree of freedom affects the momenta of the electron and hole.
Thus only the wave functions of bright excitons (both in momentum- and spin-bright) are affected by the strong light-matter coupling \cite{glazov_review,deng_tutorial}, while the wave functions of dark excitons of any kind are described by the standard Wannier equation. \gru{Calculation of the bright-exciton wave function 
\vor{with the} strong coupling between excitons and photons \vor{taken into account} is 
\vor{usually referred to} as the `flexible exciton' limit, while the `rigid exciton' limit considers Eqs.~\eqref{Wannier} 
\vor{with} the photon field \vor{disregarded}. The flexible-exciton wave function is calculated 
considering in Eqs.~\eqref{Wannier} ${\bf k}_1-{\bf k}_2=0$ and $\Omega$ 
equal to the lowest energy level which is defined by the bottom of the lower-polariton dispersion $E_{\rm LP}^0$.} Furthermore, the wave functions of different exciton species may vary slightly due to the effective masses  differences in various bands. The effects of non-equilibrium and the corresponding modifications of Eqs.~\eqref{Wannier} due to losses are discussed in Section~\ref{sec4}. 

We note that among the 16 exciton fields which could be excited, 8 fields describe $A$--excitons and the other 8 fields are related to $B$--excitons that lie higher in energy due to the presence of the energy splitting of the bands with different spins in TMDs~\cite{kormanyos}. In particular, for WS$_2$ the splitting for holes with different spins is about 400~meV \cite{ws2, ws2_chernikov}.
Overall, within the 16 exciton fields one can distinguish 4 momentum- and spin-bright, 4 momentum-bright spin-dark (referred to as SD-fields), 4 spin-bright momentum-dark (MD), and 4 that are dark both in spin and momentum (SMD).

\section{Nonlinearities: inter- and intravalley interactions}~\label{sec3}

The main goal of this paper is incorporation of valley and spin degrees of freedom to the consideration of polariton nonlinearities. On the one hand, the problem of accounting for all the electron-hole-photon coupling possibilities is combinatorial and requires treatment of all the compositions of the exciton field species; on the other hand, the inclusion of valleys results in the appearance of several distinct interaction constants.

The nonlinear term in the exciton-photon action for the spinless and valley-free case is derived in Appendix~\ref{appA} [see Eq.~\eqref{spinless_S4}]. When the spin and valley indices are restored, it becomes: 
\begin{multline}\label{nonlinear_general_initial}
    \Delta \mathcal{S}^{(4)} \!=\! -\frac{1}{4} \! \int \!  
    \frac{d\Omega_1}{2 \pi} \frac{d\Omega_2}{2 \pi} \frac{d\Omega_3}{2 \pi} \! \sum_{{\bf k}_1,\dots {\bf k}_4}\left[\varepsilon_{c \lambda_1}^{\sigma_1}({\bf k}_1) + \varepsilon_{c \lambda_3}^{\sigma_3}({\bf k}_3) \right.\\
   \qquad\qquad\qquad\qquad \left.- \varepsilon_{v \lambda_2}^{\sigma_2}({\bf k}_2)- \varepsilon_{v\lambda_4}^{\sigma_4}({\bf k}_4) +  \Omega_1 +  \Omega_3\right]\times 
   \\[-3pt] 
  \!\!\!\times  \biggl\{[\hat{\Delta}^{\sigma_3 \sigma_2}_{\lambda_3 \lambda_2}]^{\dag}({\bf k}_3, {\bf k}_2, \Omega_2) \, \hat\Delta^{\sigma_1 \sigma_2}_{\lambda_1\lambda_2}({\bf k}_1,{\bf k}_2, \Omega_1) \Biggr. \qquad\qquad\qquad  \\[-6pt]
  \Biggl.  [\hat{\Delta}^{\sigma_1\sigma_4}_{\lambda_1\lambda_4}]^{\dag}({\bf k}_1, {\bf k}_4, \Omega_1 \!+\! \Omega_3 \!-\! \Omega_2) \, \hat{\tau}_1 \, \hat\Delta^{\sigma_3\sigma_4}_{\lambda_3\lambda_4}({\bf k}_3, {\bf k}_4, \Omega_3) \\[-3pt]
+ [\hat{\Delta}^{\sigma_3 \sigma_2}_{\lambda_3 \lambda_2}]^{\dag}({\bf k}_3, {\bf k}_2, \Omega_2) \,\hat\Delta^{\sigma_3\sigma_4}_{\lambda_3\lambda_4}({\bf k}_3, {\bf k}_4, \Omega_3) \qquad\qquad \\[-3pt] 
\Biggl. [\hat{\Delta}^{\sigma_1\sigma_4}_{\lambda_1\lambda_4}]^{\dag}({\bf k}_1, {\bf k}_4, \Omega_1 \!+\! \Omega_3 \!-\! \Omega_2) \, \hat{\tau}_1 \, \hat\Delta^{\sigma_1 \sigma_2}_{\lambda_1\lambda_2}({\bf k}_1,{\bf k}_2, \Omega_1)
    \biggr\},
\end{multline}
where $\hat{\tau}_{1}$ is the Pauli matrix in the Keldysh space.
From the point of view of the valley indices, there are four different combinations that describe (i) purely intravalley 
\begin{table}[h!]
\caption{\label{table1} Classification of interactions }
\begin{ruledtabular}
\begin{tabular}{lcc}
Type of interaction & Type of interaction  \\
 &  (valley composition)\\
\hline
(i) intravalley &  $\lambda_1 = \lambda_2=\lambda_3=\lambda_4$  \\[2pt] 
 (ii) intervalley &  $\lambda_1 = \lambda_3 \neq \lambda_2 =  \lambda_4$ \\[2pt]
(iii) inter- and intravalley &  $\lambda_1 = \lambda_2 = \lambda_3 \neq \lambda_4$\footnote{and three other circlic permutations}  \\[2pt]
(iv) valley change&  $\lambda_1 \neq \lambda_3$, $\lambda_2 \neq \lambda_4$  
\end{tabular}
\end{ruledtabular}
\end{table}
interactions, (ii) purely intervalley interactions, (iii) interactions between inter- and intravalley excitons, and (iv) interactions with the valley change (Table~\ref{table1} shows the valley compositions for all types).

All the derivations of the spin- and valley-dependent interactions are cumbersome, but  rather straightforward. The Wannier equations~\eqref{Wannier} at the same time define the configuration of the exciton field $\Delta^{\sigma \sigma'}_{\lambda\lambda'}(k_1, k_2)$ 
and allow to rewrite~\eqref{nonlinear_general_initial} in the shape
\begin{widetext}
    \begin{multline}\label{nonlinear_general} 
   \Delta \mathcal{S}^{(4)} = -\frac{1}{8}\int 
    \frac{d\Omega_1}{2 \pi} \frac{d\Omega_2}{2 \pi} \frac{d\Omega_3}{2 \pi} \sum_{{\bf k}_1\dots {\bf k}_4, {\bf q}} V({\bf q})\times \\[-6pt] \times\biggl\{[\hat{\Delta}^{\sigma_3 \sigma_2}_{\lambda_3 \lambda_2}]^{\dag}({\bf k}_3-{\bf q} , {\bf k}_2 -{\bf q}, \Omega_2) \hat\Delta^{\sigma_1 \sigma_2}_{\lambda_1\lambda_2}({\bf k}_1,{\bf k}_2, \Omega_1)[\hat{\Delta}^{\sigma_1\sigma_4}_{\lambda_1\lambda_4}]^{\dag}({\bf k}_1, {\bf k}_4, \Omega_1 + \Omega_3 -\Omega_2) \hat{\tau}_1\hat\Delta^{\sigma_3\sigma_4}_{\lambda_3\lambda_4}({\bf k}_3, {\bf k}_4, \Omega_3) \\
    +[\hat{\Delta}^{\sigma_3 \sigma_2}_{\lambda_3 \lambda_2}]^{\dag}({\bf k}_3 , {\bf k}_2, \Omega_2) \hat\Delta^{\sigma_1 \sigma_2}_{\lambda_1\lambda_2}({\bf k}_1-{\bf q},{\bf k}_2-{\bf q}, \Omega_1)[\hat{\Delta}^{\sigma_1\sigma_4}_{\lambda_1\lambda_4}]^{\dag}({\bf k}_1, {\bf k}_4, \Omega_1 + \Omega_3 -\Omega_2) \hat{\tau}_1\hat\Delta^{\sigma_3\sigma_4}_{\lambda_3\lambda_4}({\bf k}_3, {\bf k}_4, \Omega_3) \\[3pt]
    +[\hat{\Delta}^{\sigma_3 \sigma_2}_{\lambda_3 \lambda_2}]^{\dag}({\bf k}_3 , {\bf k}_2, \Omega_2) \hat\Delta^{\sigma_1 \sigma_2}_{\lambda_1\lambda_2}({\bf k}_1,{\bf k}_2, \Omega_1)[\hat{\Delta}^{\sigma_1\sigma_4}_{\lambda_1\lambda_4}]^{\dag}({\bf k}_1-{\bf q}, {\bf k}_4-{\bf q}, \Omega_1 + \Omega_3 -\Omega_2) \hat{\tau}_1\hat\Delta^{\sigma_3\sigma_4}_{\lambda_3\lambda_4}({\bf k}_3, {\bf k}_4, \Omega_3) \\[3pt]
    +[\hat{\Delta}^{\sigma_3 \sigma_2}_{\lambda_3 \lambda_2}]^{\dag}({\bf k}_3 , {\bf k}_2, \Omega_2) \hat\Delta^{\sigma_1 \sigma_2}_{\lambda_1\lambda_2}({\bf k}_1,{\bf k}_2, \Omega_1)[\hat{\Delta}^{\sigma_1\sigma_4}_{\lambda_1\lambda_4}]^{\dag}({\bf k}_1, {\bf k}_4, \Omega_1 + \Omega_3 -\Omega_2) \hat{\tau}_1\hat\Delta^{\sigma_3\sigma_4}_{\lambda_3\lambda_4}({\bf k}_3-{\bf q}, {\bf k}_4-{\bf q}, \Omega_3) \\[3pt]
   + [\hat{\Delta}^{\sigma_3 \sigma_2}_{\lambda_3 \lambda_2}]^{\dag}({\bf k}_3-{\bf q}, {\bf k}_2-{\bf q}, \Omega_2) \hat\Delta^{\sigma_3\sigma_4}_{\lambda_3\lambda_4}({\bf k}_3, {\bf k}_4, \Omega_3)  
[\hat{\Delta}^{\sigma_1\sigma_4}_{\lambda_1\lambda_4}]^{\dag}({\bf k}_1, {\bf k}_4, \Omega_1 + \Omega_3 -\Omega_2) \hat{\tau}_1   \hat\Delta^{\sigma_1 \sigma_2}_{\lambda_1\lambda_2}({\bf k}_1,{\bf k}_2, \Omega_1) \\[3pt]
+[\hat{\Delta}^{\sigma_3 \sigma_2}_{\lambda_3 \lambda_2}]^{\dag}({\bf k}_3, {\bf k}_2 , \Omega_2) \hat\Delta^{\sigma_3\sigma_4}_{\lambda_3\lambda_4}({\bf k}_3-{\bf q}, {\bf k}_4-{\bf q}, \Omega_3)  
[\hat{\Delta}^{\sigma_1\sigma_4}_{\lambda_1\lambda_4}]^{\dag}({\bf k}_1, {\bf k}_4, \Omega_1 + \Omega_3 -\Omega_2) \hat{\tau}_1   \hat\Delta^{\sigma_1 \sigma_2}_{\lambda_1\lambda_2}({\bf k}_1,{\bf k}_2, \Omega_1)
\\[3pt]
+ [\hat{\Delta}^{\sigma_3 \sigma_2}_{\lambda_3 \lambda_2}]^{\dag}({\bf k}_3, {\bf k}_2 , \Omega_2) \hat\Delta^{\sigma_3\sigma_4}_{\lambda_3\lambda_4}({\bf k}_3, {\bf k}_4, \Omega_3)  
[\hat{\Delta}^{\sigma_1\sigma_4}_{\lambda_1\lambda_4}]^{\dag}({\bf k}_1-{\bf q}, {\bf k}_4-{\bf q}, \Omega_1 + \Omega_3 -\Omega_2) \hat{\tau}_1   \hat\Delta^{\sigma_1 \sigma_2}_{\lambda_1\lambda_2}({\bf k}_1,{\bf k}_2, \Omega_1)
\\[-3pt]
+ [\hat{\Delta}^{\sigma_3 \sigma_2}_{\lambda_3 \lambda_2}]^{\dag}({\bf k}_3, {\bf k}_2 , \Omega_2) \hat\Delta^{\sigma_3\sigma_4}_{\lambda_3\lambda_4}({\bf k}_3, {\bf k}_4, \Omega_3)  
[\hat{\Delta}^{\sigma_1\sigma_4}_{\lambda_1\lambda_4}]^{\dag}({\bf k}_1, {\bf k}_4, \Omega_1 + \Omega_3 -\Omega_2) \hat{\tau}_1   \hat\Delta^{\sigma_1 \sigma_2}_{\lambda_1\lambda_2}({\bf k}_1-{\bf q},{\bf k}_2-{\bf q}, \Omega_1)
    \biggr\} 
    \\[3pt]
  + \frac{1}{8}\! \int\!\!\frac{d\Omega_1}{2\pi}\!\frac{d\Omega_2}{2\pi} \!\frac{d\Omega_3}{2\pi}\! \!\!\!\sum_{{\bf k}_1\dots {\bf k}_4} \!\!\!\!g_{R} \Bigl\{\hat{\Psi}_{\rm ph} ^{\sigma_3\,\dag}\!({\bf k}_3\!-\! {\bf k}_2, \Omega_2) \hat\Delta^{\sigma_1 \sigma_2}_{\!\lambda_1\lambda_2}\!({\bf k}_1,{\bf k}_2, \Omega_1\!)[\hat{\Delta}^{\sigma_1\sigma_4}_{\!\lambda_1\lambda_4}]^{\dag}\!({\bf k}_1, {\bf k}_4, \Omega_1 + \Omega_3 -\Omega_2) \hat{\tau}_1\!\hat\Delta^{\sigma_3\sigma_4}_{\!\lambda_3\lambda_4}\!({\bf k}_3, {\bf k}_4, \Omega_3)  \delta_{\sigma_3 \sigma_2} \delta_{\lambda_3 \lambda_2}\\[-2pt]
    + [\hat{\Delta}^{\sigma_3 \sigma_2}_{\lambda_3 \lambda_2}]^{\dag}({\bf k}_3, {\bf k}_2, \Omega_2) \hat\Psi_{\rm ph}^{\sigma_1}({\bf k}_1-{\bf k}_2, \Omega_1)[\hat{\Delta}^{\sigma_1\sigma_4}_{\lambda_1\lambda_4}]^{\dag}({\bf k}_1, {\bf k}_4, \Omega_1 + \Omega_3 -\Omega_2) \hat{\tau}_1\hat\Delta^{\sigma_3\sigma_4}_{\lambda_3\lambda_4}({\bf k}_3, {\bf k}_4, \Omega_3)  \delta_{\sigma_1 \sigma_2} \delta_{\lambda_1 \lambda_2} \\[3pt] 
    + [\hat{\Delta}^{\sigma_3 \sigma_2}_{\lambda_3 \lambda_2}]^{\dag}({\bf k}_3, {\bf k}_2, \Omega_2) \hat\Delta^{\sigma_1 \sigma_2}_{\lambda_1\lambda_2}({\bf k}_1,{\bf k}_2, \Omega_1)\hat{\Psi}_{\rm ph}^{\sigma_1\,\dag}({\bf k}_1- {\bf k}_4, \Omega_1 + \Omega_3 -\Omega_2) \hat{\tau}_1\hat\Delta^{\sigma_3\sigma_4}_{\lambda_3\lambda_4}({\bf k}_3, {\bf k}_4, \Omega_3) \delta_{\sigma_1 \sigma_4} \delta_{\lambda_1 \lambda_4} \\[3pt] 
    +[\hat{\Delta}^{\sigma_3 \sigma_2}_{\lambda_3 \lambda_2}]^{\dag}({\bf k}_3, {\bf k}_2, \Omega_2) \hat\Delta^{\sigma_1 \sigma_2}_{\lambda_1\lambda_2}({\bf k}_1,{\bf k}_2, \Omega_1)[\hat{\Delta}^{\sigma_1\sigma_4}_{\lambda_1\lambda_4}]^{\dag}({\bf k}_1, {\bf k}_4, \Omega_1 + \Omega_3 -\Omega_2) \hat{\tau}_1\hat\Psi_{\rm ph}^{\sigma_3}({\bf k}_3- {\bf k}_4, \Omega_3) \delta_{\sigma_3\sigma_4} \delta_{\lambda_3\lambda_4} 
\\[3pt]
 +\hat{\Psi}_{\rm ph}^{\sigma_3\,\dag}({\bf k}_3- {\bf k}_2, \Omega_2) \hat\Delta^{\sigma_3\sigma_4}_{\lambda_3\lambda_4}({\bf k}_3, {\bf k}_4, \Omega_3)  
[\hat{\Delta}^{\sigma_1\sigma_4}_{\lambda_1\lambda_4}]^{\dag}({\bf k}_1, {\bf k}_4, \Omega_1 + \Omega_3 -\Omega_2) \hat{\tau}_1   \hat\Delta^{\sigma_1 \sigma_2}_{\lambda_1\lambda_2}({\bf k}_1,{\bf k}_2, \Omega_1)\delta_{\sigma_3\sigma_2} \delta_{\lambda_3\lambda_2} \\[3pt]
+
 [\hat{\Delta}^{\sigma_3 \sigma_2}_{\lambda_3 \lambda_2}]^{\dag}({\bf k}_3, {\bf k}_2, \Omega_2) \Psi_{\rm ph}^{\sigma_3}({\bf k}_3-{\bf k}_4, \Omega_3) [\hat{\Delta}^{\sigma_1\sigma_4}_{\lambda_1\lambda_4}]^{\dag}({\bf k}_1, {\bf k}_4, \Omega_1 + \Omega_3 -\Omega_2) \hat{\tau}_1   \hat\Delta^{\sigma_1 \sigma_2}_{\lambda_1\lambda_2}({\bf k}_1,{\bf k}_2, \Omega_1)\delta_{\sigma_3\sigma_4} \delta_{\lambda_3\lambda_4}
\\[3pt] 
+ [\hat{\Delta}^{\sigma_3 \sigma_2}_{\lambda_3 \lambda_2}]^{\dag}({\bf k}_3, {\bf k}_2, \Omega_2) \hat\Delta^{\sigma_3\sigma_4}_{\lambda_3\lambda_4}({\bf k}_3, {\bf k}_4, \Omega_3)  
\hat{\Psi}_{\rm ph}^{\sigma_1\,\dag}({\bf k}_1- {\bf k}_4, \Omega_1 + \Omega_3 -\Omega_2) \hat{\tau}_1   \hat\Delta^{\sigma_1 \sigma_2}_{\lambda_1\lambda_2}({\bf k}_1,{\bf k}_2, \Omega_1)\delta_{\sigma_1\sigma_4} \delta_{\lambda_1\lambda_4}\,\,\,\\
+ [\hat{\Delta}^{\sigma_3 \sigma_2}_{\lambda_3 \lambda_2}]^{\dag}({\bf k}_3, {\bf k}_2, \Omega_2) \hat\Delta^{\sigma_3\sigma_4}_{\lambda_3\lambda_4}({\bf k}_3, {\bf k}_4, \Omega_3)  
[\hat{\Delta}^{\sigma_1\sigma_4}_{\lambda_1\lambda_4}]^{\dag}({\bf k}_1, {\bf k}_4, \Omega_1 + \Omega_3 -\Omega_2) \hat{\tau}_1   \hat\Psi_{\rm ph}^{\sigma_1 }({\bf k}_1-{\bf k}_2, \Omega_1)\delta_{\sigma_1\sigma_2} \delta_{\lambda_1\lambda_2}
    \Bigr\}.\!
    \end{multline}
\end{widetext}
Here, the first block is the exciton interaction, while the second block is the saturation nonlinearity where the fact that only bright excitons can couple to photons is taken into account. Furthermore, the correction due to the off-diagonal electron density  $\eta_{i \lambda}^{\sigma}({\bf r}, {\bf r}^\prime, t) = \overline{\Psi}_{i \lambda}^\sigma({\bf r}, t)\Psi_{i \lambda}^\sigma({\bf r}^\prime\!, t)$ that leads to the screening of the exciton interaction, has the form (see Appendix~\ref{appA} for the spinless case):
\begin{widetext}
\begin{multline}\label{nonlinear_general_scr}
   \Delta\mathcal{S}^{(4)\prime}\!\! =\! \frac{1}{4}\int \!\!
    \frac{d\Omega_1 d\Omega_2 d\Omega_3}{(2 \pi)^3} \!\!\!\sum_{{\bf k}_1,\dots {\bf k}_4,  {\bf q}}\!\!\!V({\bf q})\Bigl\{[\hat{\Delta}^{\sigma_1 \sigma_2}_{\lambda_1 \lambda_2}]^{\dag}({\bf k}_1, {\bf k}_2, \Omega_1) \hat{\tau}_1\hat\Delta^{\sigma_3 \sigma_2}_{\lambda_3 \lambda_2}({\bf k}_3, {\bf k}_2, \Omega_2)[\hat{\Delta}^{\sigma_3 \sigma_4}_{\lambda_3 \lambda_4}]^{\dag}({\bf k}_3 - {\bf q}, {\bf k}_4, \Omega_3) \times\\[-8pt] \hspace{300pt}\times\hat\Delta^{\sigma_1 \sigma_4}_{\lambda_1 \lambda_4}({\bf k}_1 - {\bf q},{\bf k}_4, \Omega_1+ \Omega_3- \Omega_2) \\
   + [\hat{\Delta}^{\sigma_3 \sigma_2}_{\lambda_3 \lambda_2}]^{\dag}({\bf k}_3 - {\bf q}, {\bf k}_2, \Omega_2) \hat{\tau}_1\hat\Delta^{\sigma_1 \sigma_2}_{\lambda_1 \lambda_2}({\bf k}_1 - {\bf q},{\bf k}_2, \Omega_1)  [\hat{\Delta}^{\sigma_1 \sigma_4}_{\lambda_1 \lambda_4}]^{\dag}({\bf k}_1, {\bf k}_4, \Omega_1+ \Omega_3- \Omega_2) \hat\Delta^{\sigma_3 \sigma_4}_{\lambda_3 \lambda_4}({\bf k}_3, {\bf k}_4, \Omega_3)\Bigr\}\delta_{\sigma_1\sigma_3}\delta_{\lambda_1\lambda_3} \\+ 
    \Bigl\{[\hat{\Delta}^{\sigma_1\sigma_2}_{\lambda_1\lambda_2}]^{\dag}({\bf k}_1, {\bf k}_2, \Omega_1) \hat{\tau}_1\hat\Delta^{\sigma_1\sigma_4}_{\lambda_1\lambda_4}({\bf k}_1, {\bf k}_4, \Omega_1+ \Omega_3- \Omega_2) [\hat{\Delta}^{\sigma_3\sigma_4}_{\lambda_3\lambda_4}]^{\dag}({\bf k}_3,{\bf k}_4 - {\bf q}, \Omega_3) \hat\Delta^{\sigma_3\sigma_2}_{\lambda_3\lambda_2}({\bf k}_3,{\bf k}_2  - {\bf q},  \Omega_2) +  
    \\+[\hat{\Delta}^{\sigma_1\sigma_4}_{\lambda_1\lambda_4}]^{\dag}({\bf k}_1,{\bf k}_4 - {\bf q}, \Omega_1+ \Omega_3- \Omega_2) \hat{\tau}_1\hat\Delta^{\sigma_1\sigma_2}_{\lambda_1\lambda_2}({\bf k}_1,{\bf k}_2  - {\bf q}, \Omega_1) [\hat{\Delta}^{\sigma_3\sigma_2}_{\lambda_3\lambda_2}]^{\dag}({\bf k}_3, {\bf k}_2, \Omega_2) \hat\Delta^{\sigma_3\sigma_4}_{\lambda_3\lambda_4}({\bf k}_3, {\bf k}_4, \Omega_3)\Bigr\}\delta_{\sigma_2\sigma_4}\delta_{\lambda_2\lambda_4}.
\end{multline}
\end{widetext}
Eq.~\eqref{nonlinear_general_scr} is the general expression for the renormalization of the exciton interaction constant due to the electron screening, containing the contributions of the type $|\Delta^{\sigma_1 \sigma_2}_{\lambda_1\lambda_2}|^2 |\Delta^{\sigma_1 \sigma_4}_{\lambda_1 \lambda_4}|^2$ (or $|\Delta^{\sigma_1 \sigma_2}_{\lambda_1\lambda_2}|^2 |\Delta^{\sigma_3 \sigma_2}_{\lambda_3 \lambda_2}|^2$) which include the interactions between intravalley spin-bright excitons $\sigma_2=\sigma_4$, $\lambda_2=\lambda_4$ (or $\sigma_1=\sigma_3$, $\lambda_1=\lambda_3$), intravalley spin-dark excitons $\sigma_2\ne\sigma_4$, $\lambda_2=\lambda_4$ (or $\sigma_1\ne\sigma_3$, $\lambda_1=\lambda_3$), intervalley spin-bright excitons $\sigma_2=\sigma_4$, $\lambda_2\ne \lambda_4$ (or $\sigma_1=\sigma_3$, $\lambda_1\ne \lambda_3$), and intervalley spin dark exctions $\sigma_2\ne \sigma_4$, $\lambda_2\ne\lambda_4$ (or $\sigma_1\ne\sigma_3$, $\lambda_1\ne\lambda_3$).

Below, we introduce the matrix elements in a similar fashion as in the standard valley-free case for $1s$ exciton limit, keeping in mind the total ${\bf k} = {\bf k}_{1} - {\bf k}_2$ and relative ${\bf p} = (m_{c\lambda}^{\sigma} {\bf k}_1 + m_{v\lambda'}^{\sigma'} {\bf k}_2)/(m_{c\lambda}^{\sigma}+ m_{v\lambda'}^{\sigma'})$ momenta of the exciton fields are valley-dependent (i.e., the relative and total momenta contain indices corresponding to ${\bf K}$ and/or ${\bf K}'$ valleys). 
Precisely, separation of variables in Eq.~\eqref{nonlinear_general} in the $1s$ exciton limit leads to the appearance of various interaction matrix elements of the general shape 
\begin{multline}\label{matrix_el}
    V_{\rm ex}({\bf l}_1, {\bf l}_2, {\bf l}_3)\! =\!\! \int \! \! \frac{d{\bf p}}{2\pi} \frac{d{\bf q}}{2\pi} \, V({\bf q}) \, \bar\chi^{\sigma_3\sigma_2}_{\lambda_3\lambda_2}({\bf p} - {\bf q}- \alpha_{12}^1 {\bf l}_1 + \alpha_{32}^3 {\bf l}_2) \\[5pt] 
    \qquad \quad \times \chi^{\sigma_1\sigma_2}_{\lambda_1 \lambda_2}\!({\bf p})   \, \bar\chi^{\sigma_1\sigma_4}_{\lambda_1\lambda_4} [{\bf p} \!+\! \alpha_{12}^2{\bf l}_1 \!-\! \alpha_{14}^4({\bf l}_1+{\bf l}_3- {\bf l}_2)] \\[5pt] \times \chi^{\sigma_3\sigma_4}_{\lambda_3 \lambda_4}({\bf p} +{\bf l}_2- \alpha_{12}^1{\bf l}_1  - \alpha_{34}^4{\bf l}_3 ), \qquad \qquad
\end{multline}
where ${\bf l}_i$ is the total exciton momentum, and the notations $\alpha_{ab}^a = m_{c\lambda}^{\sigma_a}/(m_{c\lambda}^{\sigma_a}+ m_{v\lambda'}^{\sigma_b})$, $\alpha_{ab}^b = m_{v\lambda'}^{\sigma_b}/(m_{c\lambda}^{\sigma_a}+ m_{v\lambda'}^{\sigma_b})$ are introduced. We note that the limit of a large system size is considered.

Since the type (i) of valley composition is the `standard' interaction of momentum-bright excitons which reside in the same valley, the matrix elements for the exciton interaction and saturation (and hence the contributions to $g_{\rm ex}$ and $g_{\rm sat}$) are qualitatively similar to the expressions obtained in the spinless, valley-free case; these are considered in Appendix~\ref{appD} [in particular, the arising interaction and saturation constants are given in Eqs.~\eqref{gex_KK},~\eqref{gex11} and~\eqref{g_sat_D}].

The type (ii) is the interaction of MD excitons, i.e. those residing in different valleys, for which the matrix elements depend on the indices $\lambda$ and $\lambda'$. Due to the large total momenta ${\bf l}_i\sim {\bf K} - {\bf K}'$ (of the order of tens nm$^{-1}$), as long as the effective masses and the coefficients $\alpha_{ab}^{a/b}$ are different, the overlap integrals are reduced compared to those for intravalley interactions. If the masses of electrons and holes tend to same values (all $\alpha_{ab}^{a/b}$ are equal), the terms containing ${\bf l}_i$ vanish and one obtains the same exressions as for type (i). Hence we emphasize that the mass differences are of a great importance for intervalley interactions, as masses are controlling the overlap of the exciton wavefunctions. A similar result was obtained in Ref.~\cite{erkensten} for the interaction between different types of excitons in TMD monolayers and heterostructures. 

Interactions (iii) between intra- and intervalley excitons, as well as interactions with the valley change, are suppressed, since e.g. ${\bf l}_1$ and ${\bf l}_2$ are large and can be estimated as $\sim {\bf K-K'}$, while ${\bf l}_3$ is smaller than the characteristic momentum of the exciton wavefunction $\sim 1/a_B$. Hence the matrix element~(\ref{matrix_el}) for type (iii) is much smaller than the standard value of the interaction constant.
For exciton interactions that are accompanied by the valley change (iv), two of the total exciton momenta, e.g., ${\bf l}_1 \sim {\bf K} - {\bf K}^\prime$ and ${\bf l}_3 \sim {\bf K}^\prime - {\bf K}$ are large while the third one ${\bf l}_2$ is negligible compared to ${\bf l}_{1,3}$. Therefore the overlap integral in Eq.~(\ref{matrix_el}) is also small.

\begin{figure}[t!]
    \centering
    \includegraphics[width=0.95\linewidth]{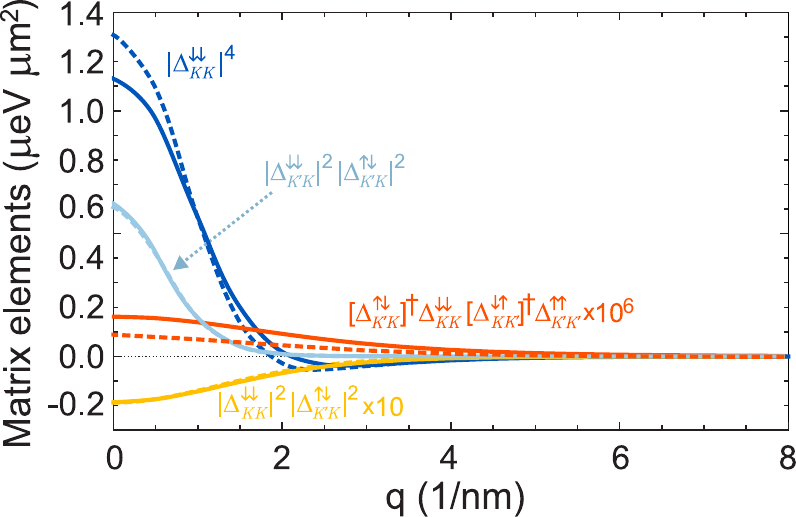}
    \caption{\small Matrix elements of exciton interaction, calculated for (i) intravalley interaction with specific case of interaction $|\Delta_{KK}^{\downarrow\downarrow}|^4$ (the dark blue lines), (ii) intervalley interaction and $|\Delta_{K'K}^{\downarrow\downarrow}|^2 |\Delta_{K'K}^{\uparrow\downarrow}|^2$ (the light blue lines), (iii) interaction between intra- and intervalley excitons for $|\Delta_{KK}^{\downarrow\downarrow}|^2 |\Delta_{K'K}^{\uparrow\downarrow}|^2$ (the yellow lines), and (iv) interaction with valley change $[\Delta_{K'K}^{\uparrow\downarrow}]^\dag \Delta_{KK}^{\downarrow\downarrow} [\Delta_{KK'}^{\downarrow\uparrow}]^\dag \Delta_{K'K'}^{\uparrow\uparrow}$ (the orange lines). \gru{The solid (dashed) lines correspond to the matrix elements calculated in the flexible (rigid) exciton limit.}  Calculations are performed for a WS$_2$ monolayer in vacuum with polarizability $\alpha_{\rm 2D} = 0.6$~nm~\cite{berkelbach} and the masses $m_{cK}^{\downarrow} = 0.27~m_0$, $m_{vK}^{\downarrow} = 0.36~m_0$, $m_{cK}^{\uparrow} = 0.36~m_0$; $a =3.18 $~\text{\AA}~\cite{kormanyos}; \gru{
    $g_R \approx 54$~meV\,nm (corresponding the Rabi splitting 
    $\hbar \Omega_R = 56$~meV), $E_{\rm ph}^0 = 1950$~meV, $E_{\rm LP}^0 = 1937$~meV~\cite{polimeno}}.
    }
    \label{fig2}
\end{figure}

In Fig.~\ref{fig2} we plot the matrix elements~(\ref{matrix_el}) with screening~(\ref{nonlinear_general_scr}) taken into account, for specific cases of exciton interactions of the types discussed above in a WS$_2$ monolayer. We perform the  calculation of the exciton wavefunction $\chi_{\lambda \lambda'}^{\sigma \sigma'}({\bf p})$ according to Eq.~(\ref{Wannier}) with the Rytova-Keldysh potential $V(q) = 2\pi e^2/[q (1 + 2\pi \alpha_{\rm 2D} q)]$, where $\alpha_{\rm 2D}$ is polarizability, \gru{for the rigid and flexible exciton cases which are shown with dashed and solid lines, respectively}. One sees that the intravalley interactions (the dark blue line) are much stronger than interactions with valley change or inter- and intravalley exction interactions (the orange and yellow lines, respectively). For the intervalley interactions (the light blue line) which are very sensitive to the effective masses ratios, in Fig.~\ref{fig2} we considered the case of interaction between two different exciton fields $\Delta_{K'K}^{\downarrow\downarrow}$ and $\Delta_{K'K}^{\uparrow\downarrow}$ (for the same field, the matrix elements will be similar to that of the type (i)). For the interaction between inter- and intravalley excitons, the resulting matrix elements are negative, which emphasizes the dominant role of hole (or electron) screening~(\ref{nonlinear_general_scr}).

\gru{It is worth noting that since \vor{we utilize} the effective mass approximation, 
the obtained exciton wave functions are 
\vor{valid} for intravalley excitons with total momenta much lower than $K\sim 1/a$, while for intervalley excitons, the effective mass approxmation is applicable for momenta comparable to $K-K'$. Thus, \vor{for large transferred momenta $q$} the \vor{obtained} exciton interaction matrix elements should be considered \vor{as a} qualitative \vor{estimate}. However, since in polariton systems the characteristic momentum is of order of several $\mu$m$^{-1}$, the effective mass approximation is 
\vor{fully applicable}; furthermore, in the relevant range of momenta, the matrix elements can be considered 
constant.} \vor{Then,}  
interaction constants are 
calculated \vor{as usual} from the matrix elements in the limit $|{\bf l}_{i}| \ll |{\bf p}|$, which is applicable for intravalley interactions (i), while for the other types (ii--iv) we consider the limit $|{\bf l}_{i} - ({\bf K} - {\bf K}^\prime)| \ll |{\bf p}|$ or  $|{\bf l}_{i} - ({\bf K}^\prime - {\bf K})| \ll |{\bf p}|$ for intervalley exciton fields. The saturation contribution that arises from the second part of Eq.~(\ref{nonlinear_general}), in the same limit of ``low'' momenta, contains 4 different saturation constants for 4 distinct saturation terms. For instance, the saturation constant corresponding to the case $\sigma_2= \sigma_3 = \sigma$, $\lambda_2 = \lambda_3 = \lambda$ is as follows: 
\begin{multline}\label{matrix_el_sat}
   g_{\rm sat}\! = g_R\!\!\int \! \! \frac{d{\bf p}}{2\pi}  \chi^{\sigma_1\sigma}_{\lambda_1\lambda}\!({\bf p})   \, \chi^{\sigma\sigma_4}_{\lambda\lambda_4}({\bf p} +{\bf l}_2- \alpha_{12}^1{\bf l}_1  - \alpha_{34}^4{\bf l}_3)
    \\[5pt]
    \times \bar\chi^{\sigma_1\sigma_4}_{\lambda_1 \lambda_4}\! ({\bf p} \!+\! \alpha_{12}^2{\bf l}_1 \!-\! \alpha_{14}^4({\bf l}_1+{\bf l}_3- {\bf l}_2)).\qquad  
\end{multline}
Here, we can distinguish several possibilities from the point of view of the valley composition: $\lambda_1 = \lambda_4 = \lambda$, which is the trivial case of intravalley (i) interactions; $\lambda_1 \neq \lambda_4 = \lambda$ corresponding to the interactions between inter- and intravalley (iii) excitons; 
$\lambda_1 = \lambda_4 \neq \lambda$ describing the photon-assisted interaction with valley change (iv). We note that the intervalley (ii) interactions are not accompanied by the saturation process. 
For the interactions of types (iii) and (iv), the
saturation constants are suppressed [compared to (i)] due to the small value of the overlap integral~(\ref{matrix_el_sat}), 
where for type (iii) ${\bf l}_{1,2}$ are small and ${\bf l}_3 \sim {\bf K} - {\bf K}^\prime$, and for type (iv) ${\bf l}_1 \sim {\bf K}^\prime  - {\bf K}$, ${\bf l}_3 \sim {\bf K}  - {\bf K}^\prime$, while ${\bf l}_2$ is low. 

\onecolumngrid

\begin{center}
\begin{table*}[h!]
\caption{\label{table2} Classification of interactions and interaction constants for the bright $A$--exciton field $\Delta_{KK}^{\downarrow\downarrow}$ in WS$_2$--based polariton system \gru{for the flexible (rigid) exciton limit}. Parameters \gru{are the same} as in Fig.~\ref{fig2}.
}
\begin{ruledtabular}
\begin{tabular}{lcccccc}
Interaction & Valley composition& Screening & Saturation & $g_{\rm ex}$~($\mu$eV$\mu$m$^{2}$) & $g_{\rm sat}$~($\mu$eV$\mu$m$^{2}$)\\
\hline
$|\Delta_{KK}^{\downarrow\downarrow}|^4$ & intravalley &  electron and hole &  4 terms & 1.1 (1.3) & 0.113 (0.18) \\[1pt] 
$|\Delta_{KK}^{\downarrow\downarrow}|^2 |\Delta_{KK}^{\uparrow\downarrow}|^2$ & intravalley &  hole &  2 terms & 6.7 (7.3) & 0.115 (0.17) \\[1pt]
$|\Delta_{KK}^{\downarrow\downarrow}|^2 |\Delta_{KK'}^{\downarrow\uparrow}|^2$ & inter- and intravalley &  electron &  2 terms & $-6.6 \,(-6.6) \times 10^{-3}$  & $2.3\, (5)\times 10^{-6}$\\[1pt]
$|\Delta_{KK}^{\downarrow\downarrow}|^2 |\Delta_{K'K}^{\downarrow\downarrow}|^2$ & inter- and intravalley &  hole &  2 terms & $-1.4\,(-1.4) \times 10^{-2}$ &  $9\, (0.9)\times 10^{-5}$\\[2pt]
$|\Delta_{KK}^{\downarrow\downarrow}|^2 |\Delta_{K'K}^{\uparrow\downarrow}|^2$ & inter- and intravalley &  hole &  2 terms & $-1.9\,(-1.8) \times 10^{-2}$ & $5.5\,(0.2)\times 10^{-6}$ \\[1pt]
$[\Delta_{KK}^{\uparrow\downarrow}]^\dag \Delta_{KK}^{\downarrow\downarrow} [\Delta_{KK'}^{\downarrow\uparrow}]^\dag \Delta_{KK'}^{\uparrow\uparrow}$ & inter- and intravalley &  no & 1 term & $4\, (1.3)\times 10^{-6}$& $2\, (3) \times 10^{-6}$ \\[1pt]
$[\Delta_{K'K}^{\downarrow\downarrow}]^\dag \Delta_{KK}^{\downarrow\downarrow} [\Delta_{KK'}^{\downarrow\uparrow}]^\dag \Delta_{K'K'}^{\downarrow\uparrow}$ & valley change &  no &  1 term & $2\, (1.1)\times 10^{-6}$  & $ 3.7 \, (5)\times 10^{-6}$ \\[1pt]
$[\Delta_{K'K}^{\uparrow\downarrow}]^\dag \Delta_{KK}^{\downarrow\downarrow} [\Delta_{KK'}^{\downarrow\uparrow}]^\dag \Delta_{K'K'}^{\uparrow\uparrow}$ & valley change &  no & \, 1 term\footnote{for circularly polarized excitation; for linearly polarized excitation 2 saturation terms arise; the values in brackets correspond to linear polarization}  & $1.6\,(0.09) \times  10^{-7}$ 
& $2.7\,(3) \times10^{-7} $ 
\end{tabular}
\end{ruledtabular}
\end{table*}
\end{center}

\begin{figure*}[t!]
    \centering \includegraphics[width= 0.96\linewidth]{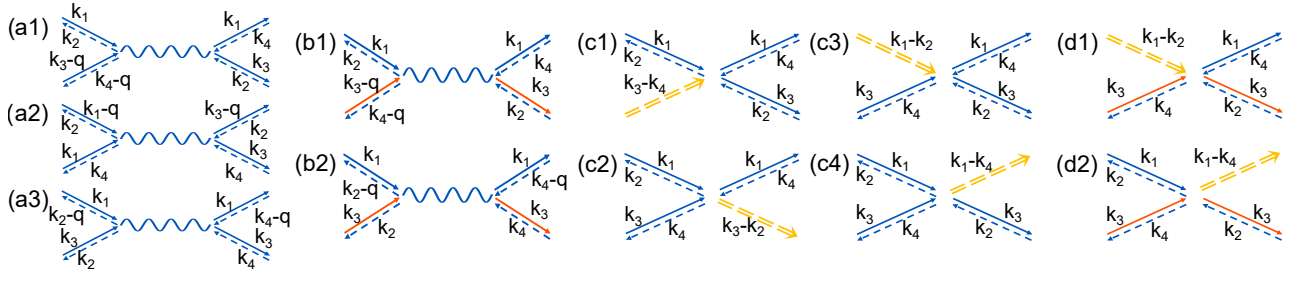} 
    \caption{\small Schematic illustration for the exciton interaction $|\Delta^{\downarrow\downarrow}_{KK}|^4$ ({a1}) and the screening contribution due to electron ({a2}) and hole ({a3}) exchange. Since $|\Delta^{\downarrow\downarrow}_{KK}|^2  |\Delta^{\uparrow\downarrow}_{KK}|^2$ ({b1}) contains two electrons in the conduction band with opposite spins, this type of interaction can be screened due to hole exchange ({b2}) only. Panels ({c1}--{c4}) and ({d1}--{d2}) show the saturation processes corresponding to nonlinearities on the panels ({a}) and ({b}), respectively. Solid and dashed lines indicate electrons in the conduction and valence bands, respectively. Blue (orange) lines correspond to electron spin-down $\downarrow$ (spin-up $\uparrow$) states. Yellow double-dashed lines show photon fields.}
    \label{fig3}
\end{figure*}
\vspace{-10pt}
\twocolumngrid

For WS$_2$, bright excitons are $\Delta^{\downarrow \downarrow}_{K K}$, $\Delta^{\uparrow \uparrow}_{K' K'}$ which are $A$--excitons, and $\Delta^{\downarrow \downarrow}_{K' K'}$, $\Delta^{\uparrow \uparrow}_{K K}$ which are $B$--excitons~\cite{comment2}.
For the sake of illustration, we will address the bright field $\Delta^{\downarrow \downarrow}_{K K}$ with the spin projection $-1$. The nonlinear processes which include this field are: interactions of $\Delta^{\downarrow \downarrow}_{K K}$; interactions between $\Delta^{\downarrow \downarrow}_{K K}$ and SD field $\Delta^{\uparrow \downarrow}_{K K}$; between $\Delta^{\downarrow \downarrow}_{K K}$ and MD field $\Delta^{\downarrow \downarrow}_{K' K}$; between $\Delta^{\downarrow \downarrow}_{K K}$ and SMD field $\Delta^{\downarrow \uparrow}_{K K'}$; there are also interactions with higher-energy excitons: with SD $\Delta^{\downarrow \uparrow}_{K K}$, MD $\Delta^{\downarrow \downarrow}_{K K'}$, and SMD $\Delta^{\uparrow \downarrow}_{K' K}$, as well as the processes accompanied by the valley change and/or spin-flip (9 various terms), i.e., altogether 16 various nonlinear contributions. Importantly, since $B$--excitons are lying much higher in energy than $A$--excitons due to the large valence band splitting in WS$_2$, in the experiments only $A$--excitons are often excited~\cite{zhao, polimeno}, so in the following we restrict ourselves with the treatment of $A$--excitons (8 terms which include interactions of types (i), (iii), and (iv)). We classify these interactions and calculate the interaction constants in Table~\ref{table2}.

Considering the main contributions to the polariton nonlinearity, 
it is important to note that for intravalley interactions of the fields $|\Delta^{\downarrow\downarrow}_{KK}|^4$, the exciton interaction constant is renormalized due to the electron and hole exchange (see Fig.~\ref{fig3}a) and the saturation nonlinearity includes 4 possible compositions for photon-mediated interactions, as shown in Fig.~\ref{fig3}c.
At the same time, for the interaction of the fields $|\Delta^{\downarrow\downarrow}_{KK}|^2|\Delta^{\uparrow\downarrow}_{KK}|^2$, the exciton interaction constant is renormalized due to the hole exchange only (see Fig.~\ref{fig3}b), since two electrons with opposite spins do not enter the screening term. This leads to the larger exciton interaction constant compared to that for the usually considered bright-exciton interaction, which underlines the important role of spin-dark excitons when estimating the polariton nonlinearities.
The saturation nonlinearity in this case is approximately two times smaller than that for the first type, since $\Delta^{\downarrow\downarrow}_{KK}$ can only convert into photons according to Eq.~(\ref{Wannier}), as is schematically illustrated in Fig.~\ref{fig3}d (for details of intravalley interactions, see Appendix~\ref{appD}).

We note that our results are applicable for the opposite polarization with the replacement $K\leftrightarrow K'$ and $\downarrow\, \leftrightarrow \, \uparrow$. It is also worth emphasizing that our theory does not include the biexciton formation (for details, see e.g. Ref.~\cite{stoof}) as well as the processes of spin and valley relaxation due to the electron-hole exchange. While for bare 2D excitons this process may lead to the relatively fast polarization decoherence~\cite{valley_coh1, valley_coh2}, in polariton systems polarization is preserved due to strong coupling~\cite{valley_pol1}.
\gru{Importantly, \vor{the presence of} strong coupling also sufficiently affects the dark exciton population. 
In WS$_2$ and WSe$_2$, the dark exciton state lies lower than \vor{the} bright one \cite{wse2_exciton}, so in exciton \vor{W-based} systems the population \vor{of dark species} is expected to be \vor{large and} long-lived~\cite{wse2_brightening, tmd_brightening} 
(
\vor{the} dark-exciton density is estimated as $\sim 10^{12}-10^{13}~$cm$^{-2}$~\cite{wse2_int, wse2_dark}).
However, in W-based {\it polariton} systems, the strong coupling may lead to the so-called brightening~\cite{polariton_brightening} when the bottom of the lower-polariton dispersion 
\vor{shifts} below the the dark-exciton energy. 
Nevertheless, dark exciton states 
\vor{are} populated due to phonon scattering~\cite{phonon_scat1, phonon_scat2}, spin relaxation~\cite{ws2_spin_relax}, or exciton-exciton Auger annihilation~\cite{auger_annihil1, auger_annihil2} which are important at room-temperature experiments. 
}

\section{Assessing nonlinearity in non-equilibrium}\label{sec4}

In the previous Sections, 
we addressed the dissipationless case $\gamma_{c,v} \to 0$ and $\gamma_{\rm ph} \to 0$. We underline that the derivation of all the expressions in the bosonization procedure presented here, such as the exciton propagator $\hat{D}({\bf k_1}, {\bf k}_2, \Omega)$, loops for the 3rd (screening terms) and 4th (nonlinear terms) orders of the expansion, even in the quasi-equilibrium limit requires the development of either the Keldysh non-equilibrium technique (detailed Appendix~\ref{appA}; see also Ref.~\cite{prl132}) or the Matsubara technique (see Ref.~\cite{prb110, stoof}). Going beyond the presented results, our current approach generally allows to treat polariton systems in nonequilibrium: the main advantage of this description is the possibility to consider losses.

When losses are non-negligible, the relation between the classical ($cl$) components of the auxiliary field $\hat{\Phi}$ and the exciton field $\hat{\Delta}$ changes and, as a result, Eqs.~(\ref{Wannier}) for the classical component of the Keldysh spinors read 
\begin{subequations}\label{Wannier_noneq}
\begin{eqnarray}
    &&\hspace{-12pt} \bigl[\varepsilon_{c\lambda}^{\sigma}({\bf k}_1) \!-\! \varepsilon_{v \lambda'}^{\sigma'}({\bf k}_2) \!-\! \Omega \!-\! i (\gamma_{c\lambda}^{\sigma} \!+\! \gamma_{v \lambda'}^{\sigma'})\bigr][\Delta_{\lambda \lambda'}^{\sigma\sigma'}]_{cl}({\bf k}_1, {\bf k}_2, \Omega) \nonumber\\
    && 
    \hspace{10pt} = \sum_{{\bf k}_3}V({\bf k}_1 \!- {\bf k}_3) [\Delta_{\lambda \lambda'}^{\sigma\sigma'}]_{cl}({\bf k}_3, {\bf k}_2 \!+ {\bf k}_3 \!-{\bf k}_1, \Omega) \nonumber
    \\[-3pt]
    &&\hspace{30pt} - \, g_{R} [\Psi_{\rm ph}^{\sigma}]_{cl}({\bf k}_1 \!- {\bf k}_2, \Omega) \delta_{\lambda\lambda'} \delta_{\sigma\sigma'} 
    \\
    &&\hspace{30pt} + \, [D_{\lambda\lambda'}^{\sigma\sigma'}]_{K}({\bf k}_1, {\bf k}_2, \Omega) [\Delta_{\lambda \lambda'}^{\sigma\sigma'}]_{q}({\bf k}_1, {\bf k}_2, \Omega) , \nonumber 
    \\[10pt]
    &&\hspace{-12pt} \bigl[\Omega - E_{\rm ph}^{\sigma}({\bf k}_1 \!- {\bf k}_2) + i \gamma_{\rm ph}^\sigma\bigr] [\Psi_{\rm ph}^{\sigma}]_{cl}({\bf k}_1 \!- {\bf k}_2, \Omega)\nonumber 
    \\
    && 
    \hspace{30pt} -\, g_{\rm R}  \sum_{{\bf k}_2} [\Delta_{\lambda\lambda}^{\sigma\sigma}]_{cl}({\bf k}_1, {\bf k}_2, \Omega) 
    \\[-3pt] 
    &&\hspace{30pt} +\, 2i \gamma_{\rm ph}^{\sigma}F_{\rm ph}(\Omega) [\Psi_{\rm ph}^{\sigma}]_q({\bf k}_1\! - {\bf k}_2, \Omega) = 0, \nonumber
\end{eqnarray}
\end{subequations}
where $\gamma_{c(v)\lambda}^{\sigma}$  are the decay rates of electrons with the spin projection $\sigma$ in conduction (valence) bands at valley $\lambda$, $\gamma_{\rm ph}^{\sigma}$ is the decay rate of photons with polarization $\sigma$, $F_{\rm ph}(\Omega) = 1 + 2 n_{\rm ph}^B(\Omega)$, $n_{\rm ph}^B(\Omega)$ is the photon bath distribution, and $[D_{\lambda\lambda'}^{\sigma\sigma'}]_{K}$ is the exciton Keldysh Green's function with restored valley and spin indices (see Appendix \ref{appA} for details). 
At the same time, the Wannier equations for the quantum $(q)$ component do not change their shape compared to Eqs.~\eqref{Wannier}. 
Due to this asymmetry between the equations which define the configurations of classical and quantum components of $\hat \Delta$, the 4th order of the expansion series cannot be derived in the same way as Eq.~(\ref{nonlinear_general_initial}) (yielding a relatively short expression). Similarly, Eq.~(\ref{nonlinear_general}) cannot be rewritten using the overlap integral with the Fourier image of the electrostatic potential $V(q)$. Therefore in the nonequilibrium case one needs to treat all the contributions to nonlinearity in the most general form that does not allow introduction of compact quantitites such as the interaction and saturation constants. This general treatment is the subject of a separate work.

\gru{As a short comment on the possible effect of losses, we consider non-zero photon 
\vor{decay} $\gamma_{\rm ph} = 1~$meV, neglecting electron and hole losses. In this case, 
correction to Eqs.~\eqref{Wannier_noneq} is of the same order of smallness as $\gamma_{\rm ph}/(E_{\rm ph}^0 - E_{\rm LP}^0)$: in the strong coupling regime, this term can be estimated as $\gamma_{\rm ph}/\hbar \Omega_R \ll1$, 
so \vor{that} the conservative expressions can be applied for the rough estimate of \vor{the} nonlinearity even for $\gamma_{\rm ph}\ne 0$. In fact, the account 
of non-zero losses leads to just a slight change of the interaction constants: e.g., for $|\Delta_{KK}^{\downarrow\downarrow}|^4$ interactions we obtain 
$g_{\rm ex} = 1.13~\mu$eV$\mu$m$^2$ and $g_{\rm sat} = 0.114~\mu$eV$\mu$m$^2$ 
(cf. the values in Table~\ref{table2}) under the assumption $F_{\rm ph}\approx1$, since the \vor{energy} of the lower-polariton dispersion bottom $E_{\rm LP}^0$ is much larger than \vor{the} temperature of thermalized photons outside the cavity (even for room $T$).}

\section{Conclusion}\label{sec5}
To summarize, we developed the Keldysh nonequilibrium theory of bosonization in an exciton-polariton system that possesses multiple different degrees of freedom. In particular, it was demonstrated that for TMD monolayer-based polaritons, all the nonlinearities can be distinguished in terms of valley compositions: four types of exciton interactions arise, leading to 16 sufficiently different contributions to nonlinearity. 
Although the main contribution is, clearly, produced by the intravalley excitons, the dominant role is not played by the bright-exciton interaction that is usually the only one considered. On the contrary, due to the difference in screening processes, for the intravalley exciton fields the interaction  between spin-dark and spin-bright species is almost an order of magnitude higher (as those are screened due to exchange of hole or electron only). This result emphasizes the crucial role of spin-dark excitons in understanding interactions in polariton systems (including other material platforms) where the dark excitons are inevitably present. Furthermore, the interactions between intervalley excitons demonstrates a peculiar susceptibility to the effective mass differences in different valleys. We also addressed the interactions between momentum-bright and momentum-dark excitons that involve valley changes (these are mostly suppressed due to the small overlap of the wavefunctions). The saturation contributions to  nonlinearity are defined by the specific choice of the four existing types of exciton interactions (intravalley, inter- and intravalley, valley change). More precisely, the saturation terms are dependent on the number of momentum- and spin-bright exciton fields participating in the process, as well as the polarization of the exciting light. 

In the existing literature, the interaction constants for WS$_2$ were initially theoretically calculated using the hydrogenic ansatz for the exciton wavefunctions~\cite{shahnazaryan2017}, which leads to unreliable numbers~\gru{\cite{ prb110, marchetti}}. The experimental values for WS$_2$ monolayer in vacuum, $g_{\rm ex} = 1.11~\mu$eV$\mu$m$^2$ and $g_{\rm sat} = 0.12~\mu$eV$\mu$m$^2$~\cite{polimeno} are in good agreement with the values obtained here. 
This agreement seemingly indicates the absence of dark exciton population, despite the nonresonant pumping used. 
However, in Ref.~\cite{polimeno} the particle density estimate was given as the upper bound, hence the experimental numbers for interaction constants provide, in fact, their lower boundary (possibly masking the presence of dark excitons). For hBN-encapsulated mono- and multilayers, the reported values of interaction constants differ drastically: $g_{\rm ex} \ll g_{\rm sat} = 10~\mu$eV$\mu$m$^2$~\cite{menon}; $g_{\rm ex}= 0.055~\mu$eV$\mu$m$^2$ and $g_{\rm sat}= 0.11~\mu$eV$\mu$m$^2$~\cite{zhao}; $g_{\rm ex}= 0.1-0.8~\mu$eV$\mu$m$^2$~\cite{landscape_confinement}. While interactions in medium are generally weaker than those in vacuum and cannot be compared with the numbers obtained in this work, such differences in reports for the same material indicate that, indeed, many different mechanisms are at play, one of which is the presence of dark excitons (together with the inevitable challenge of the accurate assessment of the polariton density). At the same time, for encapsulated TMD materials phonon-induced interactions were demonstrated to dominate the nonlinearity in the system~\cite{zhao_phonon}.

We note that this work presents the first rigorous approach to bosonization accounting for all valleys and spins of excitons, and allows for consideration of other degrees of freedom (e.g., layer indices in bilayer systems) and straightforward account for all types of nonlinearity. Importantly, the developed nonequilibrium approach naturally includes the exciton and photon losses through the introduction of the corresponding baths; when losses are nonzero, the Wannier equations for the classical and quantum components of the exciton fields start to differ. This calls for careful interpretation of the observed exciton-polariton blueshifts when the systems are strongly out of equilibrium, as a simple introduction of simplified quantities such as the interaction and saturation constants is no longer possible.

\acknowledgments{A.G. acknowledges the financial support of the BASIS Foundation under the grant No. 25-1-4-9-1. The work is supported by Ministry of Science and Higher Education of the Russian Federation (Goszadaniye) Project No. FSMG-2026-0012.}

\onecolumngrid

\appendix
\section{Derivation of the exciton-photon action in the non-equilibrium case}\label{appA}

We employ the non-equilibrium path integral approach to derive the zero-temperature exciton-photon action and the modified Wannier equation. We start with the action for the system coupled to baths: $\mathcal{S} = \mathcal{S}_{s} +  \mathcal{S}_b +\mathcal{S}_{sb} $. The first term
\begin{multline}\label{action_e-h-ph_noneq}
    \mathcal{S}_{s}[\Psi_{\!c},\Psi_{\!v}, \Psi_{\rm ph}] = \!\sum_{\lambda, \sigma} \int \!\! d{\bf r} \!\int_{\mathcal{C}} \!d t
    \left[\!\begin{pmatrix}
            \overline{\Psi}_{\!c \lambda}^{\sigma}(x) & \!\!\overline{\Psi}_{\!v\lambda}^{\sigma}(x)
        \end{pmatrix}\!\!
        \begin{pmatrix}
           i \partial_{t} \!-\! \varepsilon_{\!c \lambda}^{\sigma}({\bf\hat{k}}) & 0 \\ 0 &  \!\!\!i \partial_{t} \!-\! \varepsilon_{\!v\lambda}^{\sigma}({\bf\hat{k}}) 
        \end{pmatrix} \!\!
        \begin{pmatrix}
             {\Psi}_{\!c\lambda}^{\sigma}(x) \\ {\Psi}_{\!v\lambda}^{\sigma}(x)
        \end{pmatrix} \right.  
        +  \overline{\Psi}_{\rm ph}^{\sigma}(x)(i\partial_{t} - E_{\rm ph}({\bf\hat{k}})) \Psi_{\rm ph}^{\sigma}(x) 
        \\
        \biggl. - g_{\rm R} \left(\overline{\Psi}_{\rm ph}^{\sigma}(x)\Psi_{\!c \lambda}^{\sigma}(x)\overline{\Psi}_{\!v\lambda}^{\sigma}(x) \!
        + \!{\rm c.c.} \!\right)\!\biggr] 
        - \frac{1}{2} \!\sum\limits_{i,j}\sum\limits_{\substack{\sigma, \sigma'\\ \lambda, \lambda'}}\!\int \!\! d{\bf r}d{\bf r}^\prime \!\! \int \! dt dt^\prime V(x-x^\prime) \overline{\Psi}_{\!i \lambda}^{\sigma}(x)\Psi_{\!i \lambda}^{\sigma}(x) \overline{\Psi}_{\!j \lambda'}^{\sigma'}(x^\prime) \Psi_{\!j \lambda'}^{\sigma'}(x^\prime)
\end{multline}
describes the electron-hole-photon system defined on the Keldysh time contour. The parts of the action corresponding to the baths and the system-baths interaction are as follows:
\begin{multline}
    \mathcal{S}_{sb} + \mathcal{S}_b \!=\!\int_{\mathcal{C}} \!dt \!\sum_{\lambda, \sigma}\Biggl[ \sum_{\kappa}\begin{pmatrix}
            \overline{B}_{\!c \lambda}^{\sigma}(\kappa) & \!\!\overline{B}_{\!v\lambda}^{\sigma}(\kappa)
        \end{pmatrix}\!\!
        \begin{pmatrix}
           i \partial_{t} \!-\! \zeta_{c \lambda}^{\sigma}(\kappa) & 0 \\ 0 &  \!\!\!i \partial_{t} \!-\! \ \zeta_{v\lambda}^{\sigma}(\kappa) 
        \end{pmatrix} \!\!
        \begin{pmatrix}
             {B}_{\!c\lambda}^{\sigma}(\kappa) \\ {B}_{\!v\lambda}^{\sigma}(\kappa)
        \end{pmatrix} - \sum_{{\bf k}, \kappa} \Gamma_{i\lambda}^{\sigma}({\bf k}, \kappa)\overline{\Psi}_{i\lambda}^{\sigma}({\bf k}, t)B_{i\lambda}^{\sigma}(\kappa,t) + {\rm c.c.}\Biggr] 
        \\[-3pt] +\int_{\mathcal{C}} dt \Biggl[ \sum_{\kappa} \overline{B}_{\rm ph}^{\sigma}(\kappa) (i\partial_{t} -\zeta_{\rm ph}^{\sigma}) B_{\rm ph}^{\sigma}(\kappa)- \sum_{{\bf k}, \kappa} \Gamma_{\rm ph}^{\sigma}({\bf k}, \kappa) \overline{B}_{\rm ph}^{\sigma}(\kappa,t) \Psi_{\rm ph}^{\sigma}({\bf k},t) +{\rm c.c.} \Biggr],
\end{multline}
where $B_{i\lambda}^{\sigma}(\kappa)$ describes the fermionic bath with the dispersion law $\zeta_{i\lambda}^{\sigma}(\kappa)$ and the momentum ${\kappa}$ coupled to the electron field $\Psi_{i\lambda}^{\sigma}({\bf k})$ with the coupling constant $\Gamma_{i\lambda}^{\sigma}({\bf k},\kappa)$, while $B_{\rm ph}^\sigma(\kappa)$ is the bosonic bath coupled to photons with polarization $\sigma$ with the coupling constant $\Gamma_{\rm ph}^{\sigma}({\bf k}, \kappa)$; time is omitted for brevity.
Here we assume that each subsystem couples to its own bath.

The evolution of the system can be expressed in terms of the forward $(f)$ and backward $(b)$ branches of the time contour and then can be rotated into the classical-quantum basis~\cite{kamenev}. The quantum  $(q)$ and classical $(cl)$ components of the bosonic field $\Psi$ are defined as $\Psi_{{\it cl}, q} = (\Psi_{f} \pm \Psi_{b})/\sqrt{2}$. For the fermionic fields, the fields on the forward and backward branches are independent, so we define the ``classical'' and ``quantum'' component using the Larkin-Ovchinnikov rotation: $\Psi_{1,2} = (\Psi_{f} \pm \Psi_{b})/\sqrt{2}$, $\overline\Psi_{1,2} = (\overline{\Psi}_{f} \mp \overline{\Psi}_{b})/\sqrt{2}$. According to the procedure described in Refs.~\cite{kamenev, szymanska}, the integration over the bath degree of freedom, in the case of constant density of states, results in coupling constants $\Gamma$ being frequency independent, so that the contribution to the system's action reads 
\begin{equation}
    \Delta\mathcal{S}_{s} = \int\limits_{-\infty}^{+\infty} \!\frac{d\omega}{2\pi} \int d{\bf r} \Biggl[\sum_{\sigma}[\hat{\Psi}_{\rm ph}^{\sigma}]^\dag \begin{pmatrix}
        0 & -i \gamma_{\rm ph}^{\sigma} \\ i \gamma_{\rm ph}^{\sigma} & 2 i \gamma_{\rm ph }^{\sigma} F_{\rm ph}(\omega)
    \end{pmatrix} \hat{\Psi}_{\rm ph}^{\sigma} + \sum_{i, \sigma, \lambda}[\hat{\Psi}_{i\lambda}^{\sigma}]^{\dag}  \begin{pmatrix}
        i \gamma_{i \lambda}^{\sigma} & 2 i \gamma_{i\lambda}^{\sigma} F_{i \lambda}^{\sigma}(\omega)\\ 0& -i \gamma_{i\lambda}^{\sigma} 
    \end{pmatrix} \hat{\Psi}_{i\lambda}^{\sigma}\Biggr],
\end{equation}
where the hats indicate spinors in Keldysh space $\hat\Psi = (\Psi_{cl}, \Psi_{q})^{T}$ for bosonic fields or $\hat\Psi = (\Psi_{1}, \Psi_{2})^{T}$ for fermionic fields, as in the main text. The rates $\gamma_{\rm ph}^{\sigma}$ and $\gamma_{i\lambda}^{\sigma}$ correspond to the constant decays of the photon subsystem with polarization $\sigma$ and the electron subsystem in the $i$-th band at valley $\lambda$ with the spin projection $\sigma$, respectively, and can be expressed as $\gamma_{\rm ph}^{\sigma} = \pi (\Gamma_{\rm ph}^{\sigma})^2 N_{\rm ph}^{\sigma}= {\rm const}$ and $\gamma_{i \lambda}^{\sigma} = \pi (\Gamma_{i \lambda}^{\sigma})^2 N_{i\lambda}^{\sigma}= {\rm const}$, where $N$ is the bath's density of states. For the photon component, $F_{\rm ph}^{\sigma}(\omega) = 1 + 2 n_{\rm ph}^{B\sigma}(\omega)$, where $n^{B}(\omega)$ is the photon bath distribution function, while $F_{i \lambda}^{\sigma}(\omega) = 1 - 2 n_{i \lambda}^{F\sigma}(\omega)$, $n^{F}(\omega)$ is the electron bath distribution function.

Below, we derive the exciton-photon action within the non-equilibrium (Keldysh) formalism. 
The bosonization problem for excitons within the Green's function approach was studied in Ref.~\cite{stefanucci}. The problem was also considered in the context of the BEC-BCS crossover \cite{yamaguchi}, in the context of collective excitations spectrum and spectral characteristics of the system \cite{hanai}. 
Here we apply the formalism developed in our previous work~\cite{prb110} and introduce 16 exciton fields $\Delta^{\sigma \sigma'}_{\lambda \lambda'}(k_1, k_2) = \Psi_{c \lambda}^{\sigma}(k_1)\bar \Psi_{v \lambda'}^{\sigma'}(k_2)$ in the action~\eqref{action_e-h-ph_initial}  using the auxiliary fields $\Phi^{\sigma \sigma'}_{\lambda \lambda'}(k_1, k_2)$. 
Then, if the dissipation is taken into account, the system's action becomes: 
\begin{multline}\label{B_initial}
    \mathcal{S}= \sum_{i = c, v} \sum_{\sigma = \uparrow, \downarrow} \sum_{\lambda=K, K'}\int\frac{d \omega}{2\pi} \int d{\bf r} \Biggl[(\hat{\Psi}_{i\lambda}^{\sigma})^{\dag}  \begin{pmatrix}
        \omega - \varepsilon_{i \lambda}^{\sigma} + i \gamma_{i \lambda}^{\sigma} & 2 i \gamma_{i \lambda}^{\sigma} F_{i \lambda}^{\sigma}(\omega)\\ 0&  \omega -\varepsilon_{i\lambda}^{\sigma}  -i\gamma_{i\lambda}^{\sigma}
    \end{pmatrix} \hat{\Psi}_{i\lambda}^{\sigma}-   g_{\rm R} \biggl\{\hat{\Psi}_{\rm ph}^{\sigma\, \dag}(x) \hat{\tau}_1 \hat{\Delta}^{\sigma \sigma}_{\lambda \lambda}(x,x) + {\rm c.c.}\biggr\}\Biggr] \\
   \qquad\qquad +  \sum_{\sigma = \uparrow, \downarrow} \int\frac{d \omega}{2\pi} \int d{\bf r}\hat{\Psi}_{\rm ph }^{\sigma\, \dag} \begin{pmatrix}
        0 &\omega -E_{\rm ph}^{\sigma } -i \gamma_{\rm ph}^{\sigma} \\\omega -E_{\rm ph}^{\sigma } + i \gamma_{\rm ph}^{\sigma} & 2 i \gamma_{\rm ph}^{\sigma} F_{\rm ph}^{\sigma}(\omega)
    \end{pmatrix} \hat{\Psi}_{\rm ph}^{\sigma}
    \\ +  \sum_{\sigma,\sigma'} \sum_{\lambda, \lambda'}\frac{d \omega}{2\pi} \int d{\bf r}d{\bf r}' \Biggl[\biggl\{[\hat{\Phi}^{\sigma\sigma'}_{\lambda \lambda'}(x,x')]^{\dag} \hat{\tau}_1\hat{\Delta}^{\sigma\sigma'}_{\lambda \lambda'}(x,x') + {\rm c.c.}\biggr\} + V(x-x')[\hat{\Delta}^{\sigma\sigma'}_{\lambda \lambda'}(x,x')]^{\dag} \hat{\tau}_1 \hat{\Delta}^{\sigma\sigma'}_{\lambda \lambda'}(x,x') \\[-5pt] -\frac{1}{\sqrt{2}} \biggl([\overline{\Phi}^{\sigma \sigma'}_{\lambda \lambda'}(x,x')]_{cl} \hat{\Psi}_{v \lambda'}^{\sigma'\,\dag}(x') \hat{\Psi}_{c \lambda}^{\sigma}(x) + [\overline{\Phi}^{\sigma \sigma'}_{\lambda \lambda'}(x,x')]_{q} \hat{\Psi}_{v \lambda'}^{\sigma'\, \dag}(x')\hat{\tau}_1 \hat{\Psi}_{c\lambda}^{ \sigma}(x) + {\rm c.c.} \biggr)\Biggr],
\end{multline}
where $\hat{\tau}_1$ is the Pauli matrix in Keldysh space. The composite indices structure $\{cl/q,i,\sigma,\lambda\}$ makes the expression for the action quite cumbersome. We note however that, on the one hand, 
these numerous degrees of freedom do not change the problem qualitatively and, on the other hand, the indices in the Keldysh space are independent on valleys and spins. Hence to obtain the action used for the results presented in the main text, we drop temporarily the spin and valley indices. Then, the integration over the fermionic fields is straightforward. 

The contribution to the action is $i\,{\rm Tr}\ln{(-i \mathcal{G}^{-1})}\! =\! i\,{\rm Tr}\ln{(-i G^{-1}_0)}+ i\,{\rm Tr} \sum_{n}\!\frac1n(-1)^{n-1} \bigl(\mathcal{G}_0 \delta\mathcal{G}^{-1}\bigr)^n$,
where $\mathcal{G}$ is the matrix in the $(c,v)$ $\times$ Keldysh space
\begin{equation}
    \mathcal{G}^{-1} = \mathcal{G}_0^{-1} + \delta\mathcal{G}^{-1} = \begin{pmatrix}
        \hat{G}_c^{-1}({\bf k}_1, \omega_1) & 0 \\
        0 & \hat{G}_v^{-1}({\bf k}_1, \omega_1) 
    \end{pmatrix} \delta_{{\bf k}_1, {\bf k}_2}\delta_{\omega_1, \omega_2}
    -\frac{1}{\sqrt{2}}\begin{pmatrix} 0 &  \hat{\Phi}({\bf k_1}, {\bf k}_2, \omega_1 -\omega_2)\\ \hat{\Phi}^\dag({\bf k_2}, {\bf k}_1, \omega_2 -\omega_1) & 0
    \end{pmatrix},
\end{equation}
and the notations in the momentum-frequency representation are as follows:
\begin{equation}\nonumber
     \hat{G}_{c(v)}^{-1} = \begin{pmatrix}
        \omega - \varepsilon_{c(v)} + i \gamma_{c(v)} & 2 i \gamma_{i} F_{i}(\omega)\\ 0&   \omega - \varepsilon_{c(v)} - i \gamma_{c(v)}
    \end{pmatrix}, \qquad \hat{\Phi} = \Phi_{cl}\hat{\mathbb{I}} + \Phi_{q} \hat{\tau}_1.
\end{equation}
Since the treatment of any non-equilibrium or temperature effects is not the main goal of this work, we focus on the simplest case of the dissipationless ($\gamma_{c,v}\to +0$) system at low temperature $T\to0$ with thermalized baths. The contribution in the second order has the following form: 
\begin{equation}\label{DD}
    \Delta \mathcal{S}^{(2)} =  -\int \frac{d\Omega}{2\pi} \sum_{{\bf k_1}, {\bf k}_2}\hat{\Phi}^\dag({\bf k}_{1}, {\bf k}_{2}, \Omega)\hat{D}^{-1}({\bf k}_{1}, {\bf k}_{2}, \Omega)\hat{\Phi}({\bf k}_{1}, {\bf k}_{2}, \Omega), \quad \text{where}\, \hat{D}^{-1}=\begin{pmatrix}
        0 & D_A^{-1} \\ D_{R}^{-1} & D_{K}^{-1}
    \end{pmatrix}. 
\end{equation}
Generally, the Green's function in~\eqref{DD} has the shape
\begin{subequations}
    \begin{eqnarray}
     && \hspace{-35pt}D_{R/A}^{-1}({\bf k}_1, {\bf k}_2, \Omega) =-\frac{i}{2} \int\frac{d\omega}{2\pi} \biggl[G_c^{R/A}({\bf k}_1,  \omega) G_v^{K}({\bf k}_2,\omega - \Omega) + G_c^{K}({\bf k}_1,\omega) G_v^{A/R}( {\bf k}_2, \omega - \Omega)\biggr],\\
     &&\hspace{-35pt} D_{K}^{-1}({\bf k}_1, {\bf k}_2, \Omega) = -\frac{i}{2} \int\frac{d\omega}{2\pi} \biggl[G_c^{R}({\bf k}_1, \omega) G_v^{A}({\bf k}_2, \omega - \Omega) + G_c^{A}({\bf k}_1,\omega) G_v^{R}({\bf k}_2, \omega - \Omega)
     +  G_c^{K}( {\bf k}_1,\omega) G_v^{K}( {\bf k}_2, \omega - \Omega)\biggr].
\end{eqnarray}
\end{subequations}
We emphasize that the form of the Green's function generally depends on the bath's distribution functions. In the case which is the focus of the main text, we assume $F_{c(v)} = \pm 1$, thus obtaining $D^{-1}_{R(A)} =  1/[\varepsilon_c({\bf k_1}) - \varepsilon_v({\bf k_2}) - \Omega \mp i (\gamma_c + \gamma_v)]$. Furthermore, in the dissipationless limit $\gamma_{c, v}\to +0$, and $D^{-1}_{K}\propto \gamma_{c}, \gamma_v\to 0 $. In the general case, relevant for the discussion in Sec.~\ref{sec4}, $D^{-1}_{K}$ has the form
\begin{multline}
    D_{K}^{-1}({\bf k}_1, {\bf k}_2, \Omega) = \\
    \frac{i (\gamma_{c} + \gamma_{v})}{(\varepsilon_{c}({\bf k}_1) - \varepsilon_{v}({\bf k}_2) - \Omega)^2 + (\gamma_{c} + \gamma_{v})^2} - \frac{i \gamma_v}{(\varepsilon_{c}({\bf k}_1)- \varepsilon_{v}({\bf k}_2)-\Omega + i\gamma_{c})^2 + \gamma_{v}^2}  - \frac{i \gamma_c}{(\varepsilon_{c}({\bf k}_1)- \varepsilon_{v}({\bf k}_2)-\Omega - i\gamma_{v})^2 + \gamma_{c}^2}.
    \nonumber
\end{multline}

In the saddle-point approximation, the auxiliary field $\hat\Phi$ can be excluded, and the contribution to the exciton-photon action~\eqref{DD} reads
\begin{equation}
    \Delta \mathcal{S}^{(2)} \!=  \!-\!\!\int\!\! \frac{d\Omega}{2\pi} \!\sum_{{\bf k}_1, {\bf k}_2}\!\hat{\Delta}^\dag({\bf k}_{1}, {\bf k}_{2}, \Omega)\!
    \begin{pmatrix}
        0 &\varepsilon_c({\bf k}_1) - \varepsilon_v({\bf k}_2) - \Omega + i (\gamma_c + \gamma_v) \\
        \varepsilon_c({\bf k}_1) - \varepsilon_v({\bf k}_2) - \Omega - i (\gamma_c + \gamma_v) & 0
    \end{pmatrix}
    \!\hat{\Delta}({\bf k}_{1}, {\bf k}_{2}, \Omega),
\end{equation}
and one obtains the equation for the configuration of the exciton field:
\begin{subequations}
\begin{eqnarray}
    && -\bigl[\varepsilon_c({\bf k}_1) - \varepsilon_v({\bf k}_2) - \Omega \mp i (\gamma_c + \gamma_v)\bigr]\Delta_{cl(q)}({\bf k}_1, {\bf k}_2, \Omega)  \nonumber\\&& \hspace{100pt} + \sum_{{\bf k}_3}V({\bf k}_1 - {\bf k}_3) \Delta_{cl(q)}({\bf k}_3, {\bf k}_2+ {\bf k}_3-{\bf k}_1, \Omega)- g_{R} [\Psi_{\rm ph}]_{cl(q)}({\bf k}_1 - {\bf k}_2, \Omega)=0, \label{spinless_Wannier_a} \\[-5pt]
    &&  \bigl[\Omega - E_{\rm ph}({\bf k}_1 - {\bf k}_2)\bigr] [\Psi_{\rm ph}]_{cl(q)}({\bf k}_1 - {\bf k}_2, \Omega) - g_{\rm R} \sum_{{\bf k}_2} \Delta_{cl(q)}({\bf k}_1 - {\bf k}_2, \Omega)=0.
\end{eqnarray}
\end{subequations}
Restoring the spin and valley indices for each exciton field, one gets Eqs.~\eqref{Wannier} of the main text. 
As one can see, the equations for classical and quantum components in this case are the same, due to the fact that the Keldysh component is zero (generally, when $D_K\ne 0$, the equations for the classical and quantum components are different, see Eq.~\eqref{Wannier_noneq}).

In the same manner, we can derive the contribution from the 4th order of the logarithm expansion taking into account the relation between the auxiliary and exciton fields: 
\begin{multline}\label{nonlinear_simple}
    \Delta \mathcal{S}^{(4)} = -\frac{1}{4}\int 
    \frac{d\Omega_1}{2 \pi} \frac{d\Omega_2}{2 \pi} \frac{d\Omega_3}{2 \pi} \sum_{{\bf k}_1 \dots {\bf k}_4}[\varepsilon_{c}({\bf k}_1) + \varepsilon_{c}({\bf k}_3) - \varepsilon_{v}({\bf k}_2)- \varepsilon_{v}({\bf k}_4) -  \Omega_1 -  \Omega_3] \times \\[-2pt] \times \Bigl[\hat{\Delta}^{\dag}({\bf k}_3, {\bf k}_2, \Omega_2)\hat\Delta({\bf k}_1,{\bf k}_2, \Omega_1)\hat{\Delta}^{\dag}({\bf k}_1, {\bf k}_4, \Omega_1 + \Omega_3 -\Omega_2) \hat{\sigma}_1\hat\Delta({\bf k}_3, {\bf k}_4, \Omega_3) \\[-5pt]+ \hat{\Delta}^{\dag}({\bf k}_3, {\bf k}_2, \Omega_2)\hat\Delta({\bf k}_3,{\bf k}_4, \Omega_3)\hat{\Delta}^{\dag}({\bf k}_1, {\bf k}_4, \Omega_1 + \Omega_3 -\Omega_2) \hat{\sigma}_1\hat\Delta({\bf k}_1, {\bf k}_2, \Omega_1) \Bigr].
\end{multline}

We should note that in the above we considered the exciton pairing channel only, while in general other pairing channels should be taken into account. In particular, the off-diagonal density field $\eta_{i}({\bf r}, {\bf r}', t) = \overline{\Psi}_{i}({\bf r}, t)\Psi_{i}({\bf r}^\prime\!, t)$ ($i=c,v$) was demonstrated to change the exciton interaction term, and the diagonal density field $ \xi_{i}({\bf r}, t) = \overline{\Psi}_{i}({\bf r}, t)\Psi_{i}({\bf r}\!, t)$ should lead to the renormalization of the one-particle dispersion due to electrostatic interactions.
For clarity, when considering the effect of the density channels, we will omit the ``photon'' part the action which does not play a role here, since the photon field does not couple with the density fields.
Turning to the initial form of the action~(\ref{B_initial}), we introduce density fields using the Hubbard-Stratonovich transformation:
 \begin{multline}
      \mathcal{S}= \int\frac{d \omega}{2\pi} \int d{\bf r} \Biggl[\sum_{i} \hat{\Psi}_{i} ^\dag \begin{pmatrix}
        \omega - \varepsilon_{i} + i \gamma_{i } & 2 i \gamma_{i} F_{i}(\omega)\\ 0&  \omega -\varepsilon_{i}  -i\gamma_{i} 
    \end{pmatrix} \hat{\Psi}_{i}\Biggr] + \int\frac{d \omega}{2\pi} \int d{\bf r}d{\bf r}' \Biggl[\biggl([\hat{\Phi}(x,x')]^{\dag} \hat{\tau}_1\hat{\Delta}(x,x') + {\rm c.c.}\biggr)
    \\  + V(x-x')\hat{\Delta}^{\dag}(x,x') \hat{\tau}_1 \hat{\Delta}(x,x') -\frac{1}{\sqrt{2}} \biggl(\overline{\Phi}_{cl}(x,x') \hat{\Psi}_{v}^{\dag}(x') \hat{\Psi}_{c }(x) + \overline{\Phi}_{q}(x,x') \hat{\Psi}_{v}^{\dag}(x')\hat{\tau}_1 \hat{\Psi}_{c }(x) + {\rm c.c.} \biggr) - \frac{1}{2}V^{-1}(x-x') \hat\xi(x)\hat{\tau}_1\hat\xi(x') \\
    -\frac{1}{\sqrt{2}} \sum_{i} \biggl(\xi_{cl}(x) \hat{\Psi}^\dag_{i}(x')\hat\Psi_{i}(x') + \xi_{q}(x) \hat{\Psi}^\dag_{i}(x')\hat{\tau}_1\hat\Psi_{i}(x') + i [\eta_i(x,x') ]_{cl}\hat{\Psi}^\dag_{i}(x')\hat\Psi_{i}(x) +i [\eta_i(x,x') ]_{q}\hat{\Psi}^\dag_{i}(x')\hat{\tau}_1\hat\Psi_{i}(x) \\[-5pt] - \frac{1}{2}V^{-1}(x-x')\hat\eta_{i}(x,x')\hat{\tau}_1\hat\eta^{q}_{i}(x',x)\biggr)\Biggr],
\end{multline}
where $\xi = \xi_c + \xi_v$. As before, the hatted quantities correspond to spinors in Keldysh space.
When compared to the ``pure excitonic'' treatment, in this case the ``perturbation'' of $\mathcal{G}^{-1}$ in the expansion is as follows:
\begin{equation}
    \delta\mathcal{G}^{-1} = -\frac{1}{\sqrt{2}}\begin{pmatrix} i \hat\eta_{c}({\bf k}_2, {\bf k}_1, \omega_2-\omega_1) +  \hat\xi({\bf k}_2- {\bf k}_1, \omega_2-\omega_1) &  \hat{\Phi}({\bf k_1}, {\bf k}_2, \omega_1 -\omega_2)\\ \hat{\Phi}^\dag({\bf k_2}, {\bf k}_1, \omega_2 -\omega_1) & i \hat\eta_{v}({\bf k}_2, {\bf k}_1, \omega_2-\omega_1) +  \hat\xi({\bf k}_2- {\bf k}_1, \omega_2-\omega_1)
    \end{pmatrix}.
\end{equation}
When one performs integration over the fermion fields in the low-density limit, the contributions to the action which are related to the density fields arise in the 3rd order in the expansion only. The 1st order should vanish due to the electroneutrality of the system, while the treatment of the 2nd order of the expansion leads to the renormazilation of the Keldysh interaction $V(q)$, which is zero in the limit of $T\to0$. In the 3rd order, one obtains:
\begin{multline}
    \Delta\mathcal{S}^{(3)} =-\frac{1}{\sqrt{2}}\int\frac{d\Omega_1}{2\pi}\frac{d\Omega_2}{2\pi}\sum_{{\bf k}_1,{\bf k}_2, {\bf k}_3} \biggl[\hat\Delta^{\dag}({\bf k}_1, {\bf k}_2, \Omega_1) \hat\tau_1 \hat\Delta({\bf k}_3, {\bf k}_2, \Omega_2) (i \eta^{cl}_{c}({\bf k}_1, {\bf k}_3, \Omega_1-\Omega_2) + \xi^{cl}_{c}({\bf k}_1-{\bf k}_3, \Omega_1-\Omega_2))\\
    \hspace{-100pt}+ \hat\Delta^{\dag}({\bf k}_1, {\bf k}_2, \Omega_1) \hat\Delta({\bf k}_3, {\bf k}_2, \Omega_2) (i \eta^q_{c}({\bf k}_1, {\bf k}_3, \Omega_1-\Omega_2) + \xi^{q}_{c}({\bf k}_1-{\bf k}_3, \Omega_1-\Omega_2)){\bf k}_3 
    \\[3pt]
  +   \hat\Delta^{\dag}({\bf k}_2, {\bf k}_1, \Omega_1) \hat\tau_1 \hat\Delta({\bf k}_2, {\bf k}_3, \Omega_2) (i \eta^{cl}_{v}({\bf k}_3, {\bf k}_1, \Omega_1-\Omega_2) + \xi^{cl}_{v}({\bf k}_3-{\bf k}_1, \Omega_2-\Omega_1))\\
  +\hat\Delta^{\dag}({\bf k}_2, {\bf k}_1, \Omega_1)  \hat\Delta({\bf k}_2, {\bf k}_3, \Omega_2) (i \eta^{q}_{v}({\bf k}_3, {\bf k}_1, \Omega_1-\Omega_2) + \xi^{q}_{v}({\bf k}_3-{\bf k}_1, \Omega_2-\Omega_1))\biggr].
\end{multline}
After averaging over the density fields $\eta_{c(v)}$ and $\xi_{c(v)}$, one can obtain the correction to the exciton interaction due to the off-diagonal density fields (screening):
\begin{multline}\label{nonlinear_simple_scr}
   \Delta\mathcal{S}^{(4)\prime} = \frac{1}{2}\int 
    \frac{d\Omega_1}{2 \pi} \frac{d\Omega_2}{2 \pi} \frac{d\Omega_3}{2 \pi} \sum_{{\bf k}_1\dots {\bf k}_4,  {\bf q}} \!\! V({\bf q})\biggl[\hat{\Delta}^{\dag}({\bf k}_1, {\bf k}_2, \Omega_1) \hat{\tau}_1\hat\Delta({\bf k}_4, {\bf k}_2, \Omega_3)\hat{\Delta}^{\dag}({\bf k}_4 - {\bf q}, {\bf k}_3, \Omega_3) \hat\Delta({\bf k}_1 - {\bf q},{\bf k}_3, \Omega_1+ \Omega_3- \Omega_2) \\[-6pt] 
    + \hat{\Delta}^{\dag}({\bf k}_2, {\bf k}_4, \Omega_1) \hat{\tau}_1\hat\Delta({\bf k}_2, {\bf k}_1, \Omega_3) \hat{\Delta}^{\dag}({\bf k}_3,{\bf k}_1 - {\bf q}, \Omega_3) \hat{\Delta}({\bf k}_3,{\bf k}_4  - {\bf q},
    \Omega_{1} + \Omega_{3} - \Omega_{2}) \biggr].  
\end{multline}
It is worth noting that, given all the spin and valley indices are restored, the integration over the diagonal density results in the corrections to the Wannier equation~(\ref{Wannier}). For the rigorous derivation of all types of corrections to the self-energy, see Ref.~\cite{steinhoff}.

In the $1s$ exciton limit, the variables can be separated as $\Delta({\bf k}_1, {\bf k}_2, \Omega)\! =\! \chi\bigl(\tfrac{m_{c} {\bf k}_1 + m_{v} {\bf k}_2 }{m_{c} + m_{v}}\bigr) \tilde{\Delta}({\bf k}_1 \!-\! {\bf k}_2, \Omega)$ and, after combining Eqs.~(\ref{nonlinear_simple}) and (\ref{nonlinear_simple_scr}), the nonlinear term in the action reads:
\begin{multline}\label{spinless_S4}
    \Delta \mathcal{S}^{(4)} = -\frac{g_{\rm ex}}{4}\int 
    \frac{d\Omega_1}{2 \pi} \frac{d\Omega_2}{2 \pi} \frac{d\Omega_3}{2 \pi} \sum_{{\bf l}_1, {\bf l}_2, {\bf l}_3} \biggl[\hat{\tilde\Delta}^{\dag}({\bf l}_2, \Omega_2)\hat{\tilde\Delta}({\bf l}_1 \Omega_1)\hat{\tilde\Delta}^{\dag}({\bf l}_1 + {\bf l}_3-{\bf l}_2, \Omega_1 + \Omega_3 -\Omega_2) \hat{\tau}_1\hat{\tilde\Delta}({\bf l}_3, \Omega_3) \\[-3pt]
    + \hat{\tilde\Delta}^{\dag}({\bf l}_2, \Omega_2)\hat{\tilde\Delta}({\bf l}_3, \Omega_3)\hat{\tilde\Delta}^{\dag}({\bf l}_1, {\bf l}_4, \Omega_1 + \Omega_3 -\Omega_2) \hat{\tau}_1\hat{\tilde\Delta}({\bf l}_1 \Omega_1) \biggr] +
    \\[-3pt]
     + \frac{g_{\rm sat}}{4} \int 
    \frac{d\Omega_1}{2 \pi} \frac{d\Omega_2}{2 \pi} \frac{d\Omega_3}{2 \pi} \sum_{{\bf l}_1, {\bf l}_2, {\bf l}_3}\biggl[\hat{\tilde\Delta}^{\dag}( {\bf l}_2, \Omega_2)\hat{\tilde\Delta}({\bf l}_1, \Omega_1)[\hat{\tilde\Delta}^{\dag}({\bf l}_1, {\bf l}_3-{\bf l}_2, \Omega_1 + \Omega_3 -\Omega_2) \hat{\tau}_1\hat\Psi_{\rm ph}({\bf l}_3, \Omega_3) + {\rm c.c.}]
    \\[-3pt]  
    + [\hat{\tilde\Delta}^{\dag}({\bf l}_2, \Omega_2)\hat{\Psi}_{\rm ph}({\bf l}_1\Omega_1) + {\rm c.c.}] \hat{\tilde\Delta}^{\dag}({\bf l}_1, {\bf l}_3-{\bf l}_2, \Omega_1 + \Omega_3 -\Omega_2) \hat{\tau}_1\hat{\tilde\Delta}({\bf k}_3, \Omega_3) \biggr],
\end{multline}
where the exciton interaction and saturation constants are introduced in the usual way~\cite{yamamoto, schwendimann,  binder, glazov_bosonization, prb110} under the assumption that the c.m. momenta ${\bf l}_i$ ($i=1,2,3$) are negligibly small compared to the momentum of internal exciton motion and the system size is large,
   \begin{subequations}
        \begin{eqnarray}\label{gex}
    && g_{\rm ex}= 2 \int \! \frac{d{\bf p}}{(2\pi)^2}  \int \! \frac{d{\bf q}}{(2\pi)^2} V({\bf q})\chi({\bf p})\chi^{*}({\bf p})\chi^{*}({\bf p} -{\bf q})\bigl[\chi({\bf p} )-\chi({\bf p -q} )\bigr],\\
    && g_{\rm sat} = g_{\rm R} \int \! \frac{d{\bf p}}{(2\pi)^2}\, \chi{(\bf p)}\chi^{*}{(\bf p)}\chi{(\bf p)},
    \label{gsat}
\end{eqnarray}
   \end{subequations}
and the saturation that plays the role of the photon-induced correction to the polariton nonlinearity arises when (\ref{nonlinear_simple}) is rewritten with the use of Eq.~(\ref{spinless_Wannier_a}). When all the exciton fields are endowed with spin and valley indices, Eq.~\eqref{spinless_S4} yields Eq.~\eqref{nonlinear_general_initial} of the main text.

\section{Momentum-bright spin-dependent exciton interactions}\label{appD}

In this Appendix, we describe interactions of momentum-bright excitons arising in the same valley, when $\lambda_{i} = K$ or $K'$. Previously, it was underlined that the interaction constants are different due to the deviations of reduced masses. 

As mentioned in the main text, the $1s$-exciton approximation allows separation of variables in the exciton field
$$\Delta^{\sigma\sigma'}_{\lambda\lambda}({\bf k}_1, {\bf k}_2, \Omega) = \chi^{\sigma \sigma'}_{\lambda\lambda}\bigl(\tfrac{m_{c \lambda}^{ \sigma} {\bf k}_1 + m_{v\lambda}^{ \sigma'} {\bf k}_2 }{m_{c\lambda}^{ \sigma} + m_{v\lambda}^{ \sigma'}}\bigr) \Tilde{\Delta}^{\sigma\sigma'}_{\lambda\lambda}({\bf k}_1 - {\bf k}_2, \Omega),$$ 
where $\chi^{\sigma\sigma'}_{\lambda\lambda}({\bf p})$ is the $1s$ exciton wavefunction obtained from the Wannier equations~\eqref{Wannier} when considering a large system (without the loss of generality, $\chi$ is assumed to be real):
\begin{equation}
    \frac{\hbar^2 {\bf p}^2}{2\mu^{\sigma\sigma'}_{\lambda\lambda}}\chi^{\sigma\sigma'}_{\lambda\lambda}({\bf p}) = \int\frac{d{\bf q}}{(2\pi)^2}\biggl[V({\bf q})+ \frac{g_R^2}{E_{\rm ph}^0 - E_{LP}^0}\delta_{\sigma\sigma'}\biggr]\chi^{\sigma\sigma'}_{\lambda\lambda}({\bf p-q}),
\end{equation}
where $1/\mu^{\sigma\sigma'}_{\lambda\lambda} = \bigl(1/m_{c\lambda}^{ \sigma} + 1/m_{v\lambda}^{ \sigma'}\bigr)$. 
Then, the nonlinear contribution $\mathcal{S}^{(4)}$ given in Eq.~(\ref{nonlinear_general}) can be rewritten as
\begin{multline}
\mathcal{S}^{(4)}=-\frac{1}{4}\int 
    \frac{d\Omega_1}{2 \pi} \frac{d\Omega_2}{2 \pi} \frac{d\Omega_3}{2 \pi} \sum_{{\bf l}_1,{\bf l}_2, {\bf l}_3} [g_{\rm ex}]_{\sigma_1 \sigma_3|\sigma_2 \sigma_4}^{\lambda\lambda|\lambda\lambda}\biggl\{[\hat{\tilde\Delta}^{\sigma_{3} \sigma_{2}}_{\lambda \lambda}]^{\dag}(l_2) \hat{\tilde\Delta}^{\sigma_1 \sigma_2}_{\lambda\lambda}(l_1)[\hat{\tilde\Delta}^{\sigma_1\sigma_4}_{\lambda\lambda}]^{\dag}(l_1+l_3-l_2) \hat{\tau}_1\hat{\tilde\Delta}^{\sigma_3\sigma_4}_{\lambda\lambda}(l_3)\\+
    [\hat{\tilde\Delta}^{\sigma_3 \sigma_2}_{\lambda \lambda}]^{\dag}(l_2) \hat{\tilde\Delta}^{\sigma_3\sigma_4}_{\lambda\lambda}(l_3)  
[\hat{\tilde\Delta}^{\sigma_1\sigma_4}_{\lambda\lambda}]^{\dag}(l_1+l_3-l_2) \hat{\tau}_1   \hat{\tilde\Delta}^{\sigma_1 \sigma_2}_{\lambda\lambda}(l_1)
    \biggr\} + \mathcal{S}_{\rm sat},
\end{multline}
where $l_i \equiv ({\bf l}_i, \Omega_i)$; the last term represents the saturation contribution and will be considered below.

The exciton interaction constant $[g_{\rm ex}]_{\sigma_1 \sigma_3|\sigma_2 \sigma_4}^{\lambda\lambda| \lambda \lambda}$ is introduced in the neglection of the total exciton momentum ${\bf l}_i$ compared to the relative momentum:
\begin{multline}\label{gex_KK}
    [g_{\rm ex}]_{\sigma_1 \sigma_3|\sigma_2 \sigma_4}^{\lambda\lambda| \lambda \lambda}\!\! =\!\! \frac{1}{2}\!\int\!\! \frac{d{\bf p}}{(2\pi)^2}\!\frac{d{\bf q}}{(2\pi)^2} \!V({\bf q})\! \biggl[ \bar\chi^{\sigma_3 \sigma_2}_{\lambda\lambda}({\bf p}) \chi^{\sigma_1 \sigma_2}_{\lambda\lambda}({\bf p})\bar\chi^{\sigma_1 \sigma_4}_{\lambda\lambda}({\bf p})\chi^{\sigma_3 \sigma_4}_{\lambda\lambda}({\bf p-q}) 
    + \bar\chi^{\sigma_3 \sigma_2}_{\lambda\lambda}({\bf p}) \chi^{\sigma_1 \sigma_2}_{\lambda\lambda}({\bf p})\bar\chi^{\sigma_1 \sigma_4}_{\lambda\lambda}({\bf p-q})\chi^{\sigma_3 \sigma_4}_{\lambda\lambda}({\bf p})\\+\bar\chi^{\sigma_3 \sigma_2}_{\lambda_3\lambda_2}({\bf p}) \chi^{\sigma_1 \sigma_2}_{\lambda_1\lambda_2}({\bf p-q})\bar\chi^{\sigma_1 \sigma_4}_{\lambda_1\lambda_4}({\bf p})\chi^{\sigma_3 \sigma_4}_{\lambda_3\lambda_4}({\bf p}) +\bar\chi^{\sigma_3 \sigma_2}_{\lambda_3\lambda_2}({\bf p-q}) \chi^{\sigma_1 \sigma_2}_{\lambda_1\lambda_2}({\bf p})\bar\chi^{\sigma_1 \sigma_4}_{\lambda_1\lambda_4}({\bf p})\chi^{\sigma_3 \sigma_4}_{\lambda_3\lambda_4}({\bf p})\biggr].
\end{multline}
The screening term $\mathcal{S}^{(4)'}$ [see Eq.~(\ref{nonlinear_general_scr})] turns into
\begin{multline}
    \mathcal{S}^{(4)'}=\frac{1}{4}\int 
    \frac{d\Omega_1}{2 \pi} \frac{d\Omega_2}{2 \pi} \frac{d\Omega_3}{2 \pi} \sum_{{\bf l}_1,{\bf l}_2, {\bf l}_3} 
  [g^{h}_{\rm ex}]_{\sigma_1 \sigma_1|\sigma_2 \sigma_4}^{\lambda\lambda| \lambda \lambda}
  \biggl\{[\hat{\tilde\Delta}^{\sigma_1 \sigma_2}_{\lambda\lambda}]^{\dag}(l_1 + l_3-l_2)\hat{\tau}_1 \hat{\tilde\Delta}^{\sigma_1 \sigma_2}_{\lambda\lambda}(l_3)[\hat{\tilde\Delta}^{\sigma_1\sigma_4}_{\lambda\lambda}]^{\dag}(l_2) \hat{\tilde\Delta}^{\sigma_1\sigma_4}_{\lambda\lambda}(l_1 )
    \biggr\} \delta_{\sigma_1\sigma_3}
    \\
   + [g^{e}_{\rm ex}]_{\sigma_1 \sigma_3|\sigma_2 \sigma_2}^{\lambda\lambda|\lambda\lambda} \biggl\{ 
    [\hat{\tilde\Delta}^{\sigma_1 \sigma_2}_{\lambda\lambda}]^{\dag}(l_1 + l_3-l_2) \hat{\tau}_1 \hat{\tilde\Delta}^{\sigma_1\sigma_2}_{\lambda\lambda}(l_1)  
[\hat{\tilde\Delta}^{\sigma_3\sigma_2}_{\lambda\lambda}]^{\dag}(l_2)  \hat{\tilde\Delta}^{\sigma_3 \sigma_2}_{\lambda\lambda}(l_3)
    \biggr\} 
    \delta_{\sigma_2 \sigma_4},
    \end{multline}
where the first term can be interpreted as hole exchange and the second one as the electron exchange, with the notations
\begin{subequations}\label{gex11} 
    \begin{eqnarray} 
  &&[g^{h}_{\rm ex}]_{\sigma_1 \sigma_1|\sigma_2 \sigma_4}^{\lambda\lambda|\lambda\lambda}=2  \int \frac{d{\bf p}}{(2\pi)^2}\frac{d{\bf q}}{(2\pi)^2} V({\bf q}) |\chi^{\sigma_1 \sigma_2}_{\lambda\lambda}({\bf p})|^2 |\chi^{\sigma_1 \sigma_4}_{\lambda\lambda}({\bf p-q})|^2,\label{gex_scrh}\\
  &&[g^{e}_{\rm ex}]_{\sigma_1 \sigma_3|\sigma_2 \sigma_2}^{\lambda\lambda|\lambda\lambda}= 2 \int \frac{d{\bf p}}{(2\pi)^2}\frac{d{\bf q}}{(2\pi)^2} V({\bf q}) |\chi^{\sigma_1 \sigma_2}_{\lambda\lambda}({\bf p})|^2 |\chi^{\sigma_3 \sigma_2}_{\lambda\lambda}({\bf p-q})|^2 \label{gex_scre}.
\end{eqnarray}
\end{subequations}
These expressions are similar to the standard formula in the absence of spins and valleys (\ref{gex}), while we note that in Eqs.~(\ref{gex_KK}, \ref{gex11}) the mass differences and, consequently, the differences of wavefunctions are taken into account.

Finally, we discuss the shape of the saturation term. In contrast to the exciton interaction, the introduction of the one constant characterizing saturation is not possible since for every species of saturation corresponding different spin compositions, interaction constants will vary as the overlaps of exciton wavefunctions will alter from each other.
The complete expression describing the saturation reads:
\begin{multline}    S_{\rm sat} \!= \! \frac{1}{8} \!\!\int\!\!\frac{d\Omega_1}{2\pi}\!\frac{d\Omega_2}{2\pi} \!\frac{d\Omega_3}{2\pi}\!\! \sum_{{\bf l}_1, {\bf l}_2, {\bf l}_3} \!\Biggl[ 
[g_{\rm sat}]^{\lambda \lambda |\lambda \lambda}_{\sigma_1\sigma_3|\sigma_3 \sigma_4}
\biggl\{ \hat{\Psi}_{\rm ph}^{\sigma_3\,\dag}(l_2) \hat{\tilde\Delta}^{\sigma_1 \sigma_3}_{\lambda\lambda}(l_1)[\hat{\tilde \Delta}^{\sigma_1 \sigma_4}_{\lambda\lambda}]^{\dag}(l_1+l_3-l_2) \hat{\tau}_1\hat{\tilde\Delta}^{\sigma_3 \sigma_4}_{\lambda\lambda}(l_3)
\\[-8pt]
 \hspace{275pt} +\hat{\Psi}_{\rm ph}^{\sigma_3\,\dag}(l_2) \hat{\tilde\Delta}^{\sigma_3\sigma_4}_{\lambda\lambda}(l_3)  
[\hat{\tilde\Delta}^{\sigma_1\sigma_4}_{\lambda\lambda}]^{\dag}(l_1+ l_3-l_2) \hat{\tau}_1   \hat{\tilde\Delta}^{\sigma_1 \sigma_3}_{\lambda\lambda}(l_1)\biggr\}
\\ + [g_{\rm sat}]^{\lambda\lambda|\lambda\lambda }_{\sigma_1\sigma_3|\sigma_1\sigma_4} \!\biggl\{\!
[\hat{\tilde\Delta}^{\sigma_3 \sigma_1}_{\lambda \lambda}]^{\dag}\!(l_2) \hat{\tilde\Delta}^{\sigma_3\sigma_4}_{\lambda\lambda}(l_3)  
[\hat{\tilde\Delta}^{\sigma_1\sigma_4}_{\lambda\lambda}]^{\dag}\!(l_1+ l_3-l_2) \hat{\tau}_1   \hat\Psi_{\rm ph}^{\sigma_1}(l_1)+ 
[\hat{\tilde\Delta}^{\sigma_3 \sigma_1}_{\lambda \lambda}]^{\dag}\!(l_2) \hat\Psi_{\rm ph}^{\sigma_1}(l_1)[\hat{\tilde\Delta}^{\sigma_1\sigma_4}_{\lambda\lambda}]^{\dag}\!(l_1+l_3-l_2) \hat{\tau}_1 \hat{\tilde\Delta}^{\sigma_3\sigma_4}_{\lambda\lambda}(l_3) \!\biggr\}
\\+
[g_{\rm sat}]^{\lambda\lambda|\lambda \lambda}_{\sigma_1\sigma_3|\sigma_2\sigma_1} \!\biggl\{\!
[\hat{\tilde\Delta}^{\sigma_3 \sigma_2}_{\lambda \lambda}]^{\dag}(l_2) \hat{\tilde\Delta}^{\sigma_1 \sigma_2}_{\lambda\lambda}(l_1)\hat{\Psi}_{\rm ph}^{\sigma_1 \, \dag}(l_1 + l_3 - l_2) \hat{\tau}_1 \hat{\tilde\Delta}^{\sigma_3\sigma_1}_{\lambda\lambda}(l_3) +
[\hat{\tilde\Delta}^{\sigma_3 \sigma_2}_{\lambda \lambda}]^{\dag}(l_2) \hat\Delta^{\sigma_3\sigma_1}_{\lambda\lambda}(l_3)  
\hat{\Psi}_{\rm ph}^{\sigma_1\, \dag}(l_1 + l_3-l_2) \hat{\tau}_1   \hat{\tilde\Delta}^{\sigma_1 \sigma_2}_{\lambda\lambda}(l_1)\! \biggr\}
\\ +
[g_{\rm sat}]^{\lambda\lambda|\lambda \lambda}_{\sigma_1\sigma_3|\sigma_2\sigma_3} \!\biggl\{\![\hat{\tilde\Delta}^{\sigma_3 \sigma_2}_{\lambda \lambda}]^{\dag}\!(l_2) \hat{\tilde\Delta}^{\sigma_1 \sigma_2}_{\lambda\lambda}(l_1)[\hat{\tilde\Delta}^{\sigma_1\sigma_3}_{\lambda\lambda}]^{\dag}\!(l_1 + l_3-l_2) \hat{\tau}_1\hat\Psi_{\rm ph}^{\sigma_3}\!(l_3) 
+ [\hat{\tilde\Delta}^{\sigma_3 \sigma_2}_{\lambda \lambda}]^{\dag}\!(l_2) \Psi_{\rm ph}^{\sigma_3}\!(l_3)  
[\hat{\tilde\Delta}^{\sigma_1\sigma_3}_{\lambda\lambda}]^{\dag}\!(l_1+l_3-l_2) \hat{\tau}_1   \hat{\tilde\Delta}^{\sigma_1 \sigma_2}_{\lambda\lambda}\!(l_1)\!\biggr\}\!
    \Biggr]
\end{multline}
with the spin-composition dependent saturation interaction constants 
\begin{subequations}\label{g_sat_D}
    \begin{eqnarray}
        & &[g_{\rm sat}]^{\lambda\lambda|\lambda\lambda}_{\sigma_1 \sigma_3|\sigma_3\sigma_4} = g_R\int\frac{d{\bf p}}{(2\pi)^2} \chi_{\lambda\lambda}^{\sigma_1\sigma_3}({\bf p}) \bar\chi_{\lambda\lambda}^{\sigma_1\sigma_4}({\bf p})\chi_{\lambda\lambda}^{\sigma_3\sigma_4}({\bf p}),\\
        & & [g_{\rm sat}]^{\lambda\lambda|\lambda\lambda}_{\sigma_1 \sigma_3|\sigma_1\sigma_4} = g_R\int\frac{d{\bf p}}{(2\pi)^2} \bar\chi_{\lambda\lambda}^{\sigma_3\sigma_1}({\bf p}) \bar\chi_{\lambda\lambda}^{\sigma_1\sigma_4}({\bf p})\chi_{\lambda\lambda}^{\sigma_3\sigma_4}({\bf p}),\\
        & & [g_{\rm sat}]^{\lambda\lambda|\lambda\lambda}_{\sigma_1 \sigma_3|\sigma_2\sigma_1} = g_R\int\frac{d{\bf p}}{(2\pi)^2} \bar\chi_{\lambda\lambda}^{\sigma_3\sigma_1}({\bf p})  \chi_{\lambda\lambda}^{\sigma_1\sigma_2}({\bf p})\chi_{\lambda\lambda}^{\sigma_3\sigma_1}({\bf p}),\\
        & & [g_{\rm sat}]^{\lambda\lambda|\lambda\lambda}_{\sigma_1 \sigma_3|\sigma_2\sigma_3} = g_R\int\frac{d{\bf p}}{(2\pi)^2} \bar\chi_{\lambda\lambda}^{\sigma_3\sigma_2}({\bf p}) \bar\chi_{\lambda\lambda}^{\sigma_1\sigma_2}({\bf p})\bar \chi_{\lambda\lambda}^{\sigma_1\sigma_3}({\bf p}).
    \end{eqnarray}
\end{subequations}
\twocolumngrid


\begin{thebibliography}{00}
\bibitem{comment1} Since the number of excitations in the system is conserved, i.e., $[H, N_{\rm tot}]=0$, where $N_{\rm tot} = N_{\rm ph} + \frac{1}{2} \bigl(N_{c}-N_{v}\bigr)$, one can redefine the Hamiltonian of the system: $H - \mu N_{\rm tot} \to H$ where the introduced parameter $\mu$ works as the ``chemical potential''~\cite{yamaguchi, hanai}.
\bibitem{comment2} Such a compositions are characteristic for WX$_2$ materials; for MoX$_2$, one needs to replace $K\leftrightarrow K'$.
\bibitem{comment3} Similar problem was considered in Ref.~\cite{prl132} but for the bilayer system with tunneling between layers, with the focus on indirect exctions.
\end{thebibliography}
\end{document}